\documentclass[journal=jacsat,manuscript=article]{achemso}

\usepackage{graphicx}
\usepackage{dcolumn}
\usepackage{bm}
\usepackage{amsmath,amssymb}
\usepackage{hyperref}
\usepackage{chemformula} 
\usepackage[T1]{fontenc} 

\SectionNumbersOn

\author{Luke C. Ugwuoke}
\affiliation[University of Pretoria]
{Department of Physics, University of Pretoria, \\Private bag X20, Hatfield 0028, South Africa}
\author{Tjaart P. J. Kr\"{u}ger}
\alsoaffiliation{Department of Physics, University of Pretoria, \\Private bag X20, Hatfield 0028, South Africa}
\email{tjaart.kruger@up.ac.za}
\phone{+27 12 420 4778}
\fax{+27 12 362 5288}

\title[Plasmon resonances in multilayer Fanoshells]
{Plasmon resonances in multilayer Fanoshells} 

\abbreviations{CNS, MNS, LSPR}

\begin{document}
	
\begin{tocentry}
\centering
\includegraphics[height = 3.5 cm, width = 7 cm]{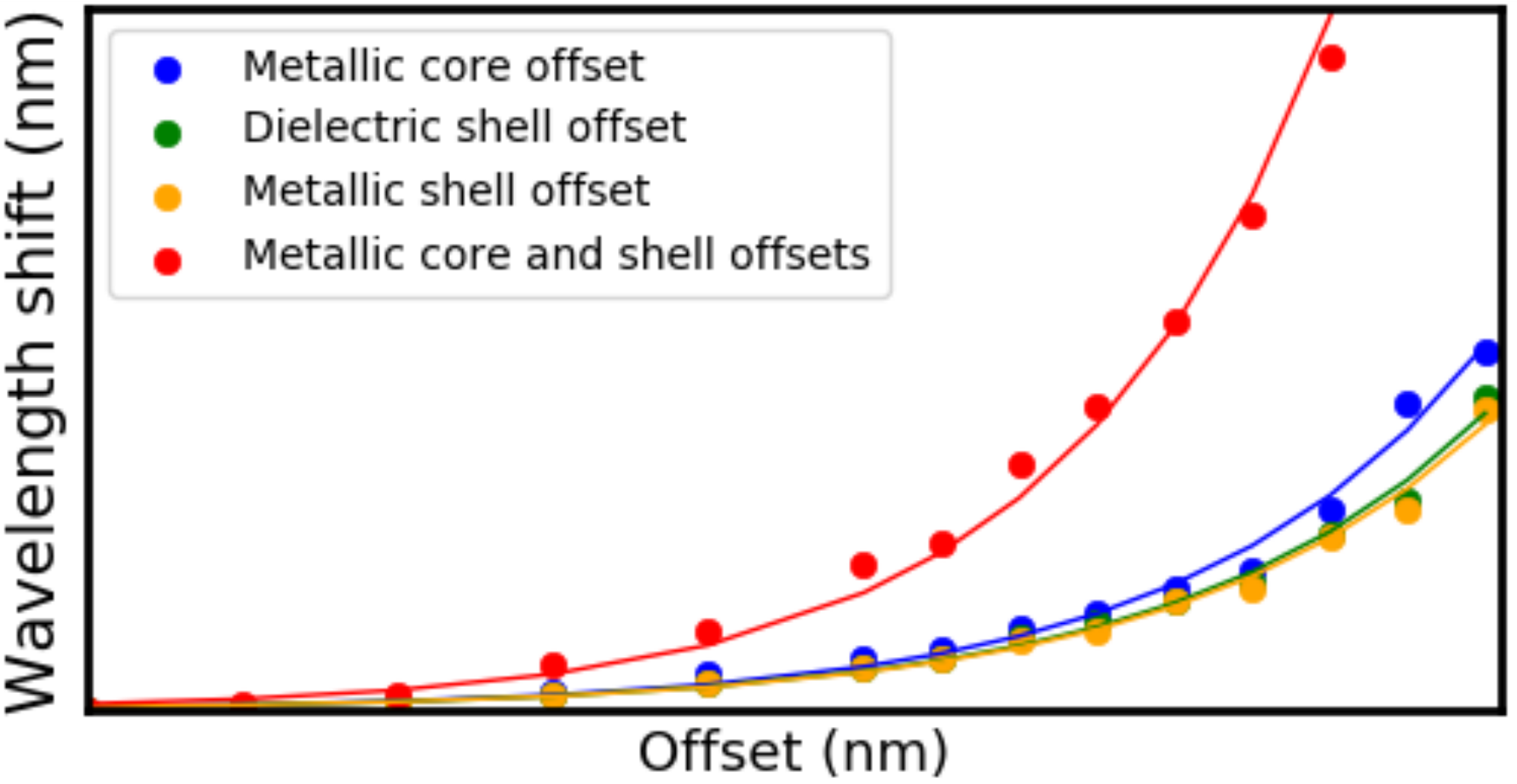}\\
\end{tocentry}
	
\begin{abstract}
	We develop a theoretical framework, based on a multipole, quasi-static approach, for the 
	prediction of the localized surface plasmon resonances in Fanoshells formed via geometrical symmetry-breaking in multilayer nanoshells consisting of a metallic core, a dielectric inner shell, and a metallic outer shell.  
	By tuning the core and shell offsets of a gold-silica-gold multilayer nanoshell, we show that the theoretical model is in good agreement with electrodynamic simulations.
	The dipolar resonances are more suppressed when the core and the outer shell are concurrently offset, and less suppressed when either the core, the inner shell, or the outer shell is offset. We attribute the former to coupling constants arising from dual symmetry-breaking, and the latter to coupling constants due to single symmetry-breaking. Using three performance parameters, we propose the outer shell offset as the optimal Fanoshell for sensing applications. 
	This study systematically investigates all types of offset-based, symmetry-broken, metal-dielectric-metal multilayer nanoshells within the Rayleigh regime.
\end{abstract}

\section{Introduction}\label{s1}
The term \emph{Fanoshells} was first used in Ref. \cite{Hala10} to refer to core-shell nanostructures capable of supporting localized surface plasmon resonances (LSPRs) with Fano-like lineshapes --- LSPRs with a distortion in their spectral lineshapes that results in an LSPR shift \cite{Hala10,Nort16,Sanc14}. 
These resonances, known as \emph{Fano resonances}, are formed as a result of interference between radiant, broad-band, dipolar modes and optically dark, narrow-band, multipolar modes such as quadrupole modes \cite{Wang06,Nort16,Qian15,Khan11}. However, the term 
\textquotedblleft Fano resonances\textquotedblright~has also been used to describe the plasmon resonances formed via the plasmon hybridization of dipolar solid sphere and nanoshell plasmons in multilayer nanoshells (MNSs) \cite{Kaj21,Hala10}.
Here, we will adopt both definitions, the latter being more relevant when there is no geometrical symmetry-breaking \cite{Sanc14,Hala10,Kaj21}, 
while the former is more relevant to symmetry-broken conventional nanoshells (CNSs) \cite{Wu06,Nort16} and MNSs \cite{Khan11,Hala10,Hu10}. In these geometries, the Fano effect has been shown to occur via dipole-quadrupole and higher-order couplings, which cause the multipolar mode to become dipole-active, i.e., the LSPR of the dark, multipolar mode becomes enhanced and visible in the spectra, while the LSPR of the bright, dipolar mode becomes less radiant or suppressed \cite{Nort16,Hala10,Khan11}. 
Mode enhancement and suppression can be revealed by calculating the dipole polarizability of the Fanoshell \cite{Nort16,Luke20} or the complex amplitude of the radiant, dipolar mode \cite{Hala10,Nort16} in order to show that such dipole-active modes are only present in the absorption and scattering spectra of symmetry-broken CNSs \cite{Nort16,Night08,Wu06} and MNSs. \cite{Hala10,Hu10}. 

Fanoshells in the sub-100 nm, Rayleigh regime are experimentally realistic using wet chemistry techniques such as those reported in Refs. \cite{Wang06,Man06}. Furthermore, the extinction spectra of CNSs excited with specially-designed laser beams also support Fano resonances \cite{Sanc14}. 
Fanoshells have been designed via geometrical symmetry-breaking in both CNSs \cite{Wu06,ZaZa13} and MNSs \cite{Qian15,Hu10}, and through the use of bimetallic nanoshells \cite{Pal11}.
Geometrical symmetry-breaking involves the introduction of some asymmetry in the arrangement of the core-shell geometry through core-offsetting \cite{Nort16,Wu06,Hala10,Hu10,Ho10}, shell-offsetting \cite{Qian15,Khan11}, and other means, such as shell-cutting \cite{Sun14}. 
This allows solid plasmons to hybridize with cavity plasmons of both the same and different orbital angular momentum numbers --- a phenomenon that is symmetry-forbidden in concentric nanoshells \cite{Wang06,Wu06,Man06}. Such an approach has led to the formation of nanoeggs from CNSs \cite{Wang06,Wu06,Night08,ZaZa13,Nort16}, a nanosphere-in-a-nanoegg from an MNS (i.e., an MNS with a metallic shell-offset \cite{Qian15,Khan11}), as well as other Fanoshell configurations such as an MNS with a metallic core-offset \cite{Hala10,Hu10}, and an MNS with a dielectric shell-offset \cite{Khan11}. 

In a CNS, solid sphere plasmons of the shell hybridize with cavity sphere plasmons of the core to form bonding and anti-bonding modes \cite{Prodan04}. 
The dipole moment of the anti-bonding mode is weak \cite{Prodan04,Gong19,Far21}, so that the subradiant, dipole-active modes in the Fanoshell of a CNS are mostly due to a Fano effect between the bonding dipole and multipolar modes \cite{Nort16,Wang06,Wu06,Night08}. 
In contrast, in an MNS, nanoshell plasmons hybridize with solid sphere plasmons of the core to form either two hybridized modes --- the bonding and anti-bonding modes, such as in Au-silica-Au and Ag-silica-Au MNSs \cite{Bard10,Zhu11,Pal11,Herr16,Nia14,Qian15,Man06} --- or three hybridized modes --- the bonding, anti-bonding and non-bonding modes, such as in Au-silica-Ag and Ag-silica-Ag MNSs \cite{Zhao12,Herr16,Nia14}. 
The dipole moment of the non-bonding mode is weak, so that it hardly undergoes any spectral shift \cite{Herr16,Nia14,Soni14,Zhao12}. However, the other two modes can be significantly enhanced, suppressed, or shifted by proper choice of geometry size and material composition. This enables MNSs to support more than one bright mode \cite{Jun14,Moud14,Hu08,Soni14}.
In a Fanoshell of an MNS, multiple LSPRs due to a Fano effect between the bonding (or anti-bonding) dipole and multipolar modes are therefore possible. \cite{Hu10,Khan11}. 

The optical cross-sections of Fano-based nanostructures feature LSPRs with linewidths usually narrower than those in the optical cross-sections of the parent nanostructures. Their narrow linewidths are very useful for refractive index sensing since they increase the figure-of-merit of plasmonic sensors \cite{Lee14,Gong19,Zang21,Li19,Far21,Ma21,Lee21}. In addition, Fano resonances are highly sensitive to changes in the refractive index of the surrounding medium \cite{Gong19,Zang21,Soni14}, undergoing LSPR shifts that are usually in accordance with the resonance shifts in the dipolar modes of the parent nanostructure. Since Fano resonances support multiple subradiant modes, certain nanostructures have been recommended for bioimaging via label-free nanolithography and surface-enhanced spectroscopy techniques \cite{Yani11}.

In Fanoshells of a CNS (i.e., nanoeggs \cite{Nort16,Wu06,Luke20,ZaZa13}), an increase in the core offset causes an enhancement of the multipolar LSPRs and a suppression of the dipolar LSPR. Likewise, in Fanoshells of an MNS, multipolar LSPRs can also be enhanced via core-offsetting but the MNS geometry creates room for additional asymmetries such as inner and outer shell offsets, which are also effective for enhancing multipolar LSPRs \cite{Khan11,Qian15}.
Previous studies have considered Fanoshells based on a Au-silica-Au MNS with either an offset Au core \cite{Hu10,Sun14,Hala10} or an offset Au shell \cite{Qian15}, a Ag-silica-Ag MNS with an offset Ag core \cite{Ho10}, or a Ag-silica-Au MNS with an offset Ag core, an offset silica shell, or an offset Au shell \cite{Khan11}. 
The core and shell offsets break the symmetry of the MNS geometry once. This is referred to as \emph{single symmetry-breaking} \cite{Hu10,Khan11}. Apart from Ref.\cite{Hala10}, which proposed a Fano oscillator model to predict the LSPRs in a Fanoshell with a Au core offset, only simulations have as yet been used to predict the steady-state extinction spectra of the single symmetry-broken Fanoshells.

Here, we employ both simulation and theoretical approaches to predict the spectra and the coupling constants responsible for single symmetry-breaking. 
In addition, we introduce a Fanoshell that breaks the symmetry of the MNS geometry twice via concurrent core and outer shell offsets, and obtain analytically the coupling constants responsible for \emph{dual symmetry-breaking} and the formation of its LSPRs. For the first time, a systematic study is presented using a generalized matrix equation applicable to metal-dielectric-metal multilayer Fanoshells of different material composition, and geometry sizes within the Rayleigh regime. 
Finally, we used three performance parameters --- geometrical sensitivity, spectral sensitivity, and scattering-to-absorption ratio --- to predict that the outer shell offset is the optimal Fanoshell for sensing applications.

\section{Theory}\label{s2}
In this work, we employ a theoretical approach based on the quasi-static approximation of Maxwell's equations \cite{Ma21,Sam07,Neev89} and the solid-harmonic addition theorem (SHAT) \cite{Nort16,Cala78}, as well as electrodynamic simulations based on three-dimensional COMSOL Multiphysics$^{\circledR}$ software \cite{Comsol}, to investigate the mode enhancement and suppression phenomena in the extinction spectra of the Fanoshells. 
In Refs. \cite{Hu10,Ho10}, where offset cores were studied, it was shown that both $x$- and $z$-polarized incident fields lead to the same number of LSPRs in the spectra of the Fanoshell but that the latter results in slightly larger spectral shifts. Therefore, we will consider only the optical response of the Fanoshells to a $z$-polarized incident field in the direction of the offsets.
\begin{figure}[ht!]
	\centering 
	\includegraphics[width = .5\textwidth]{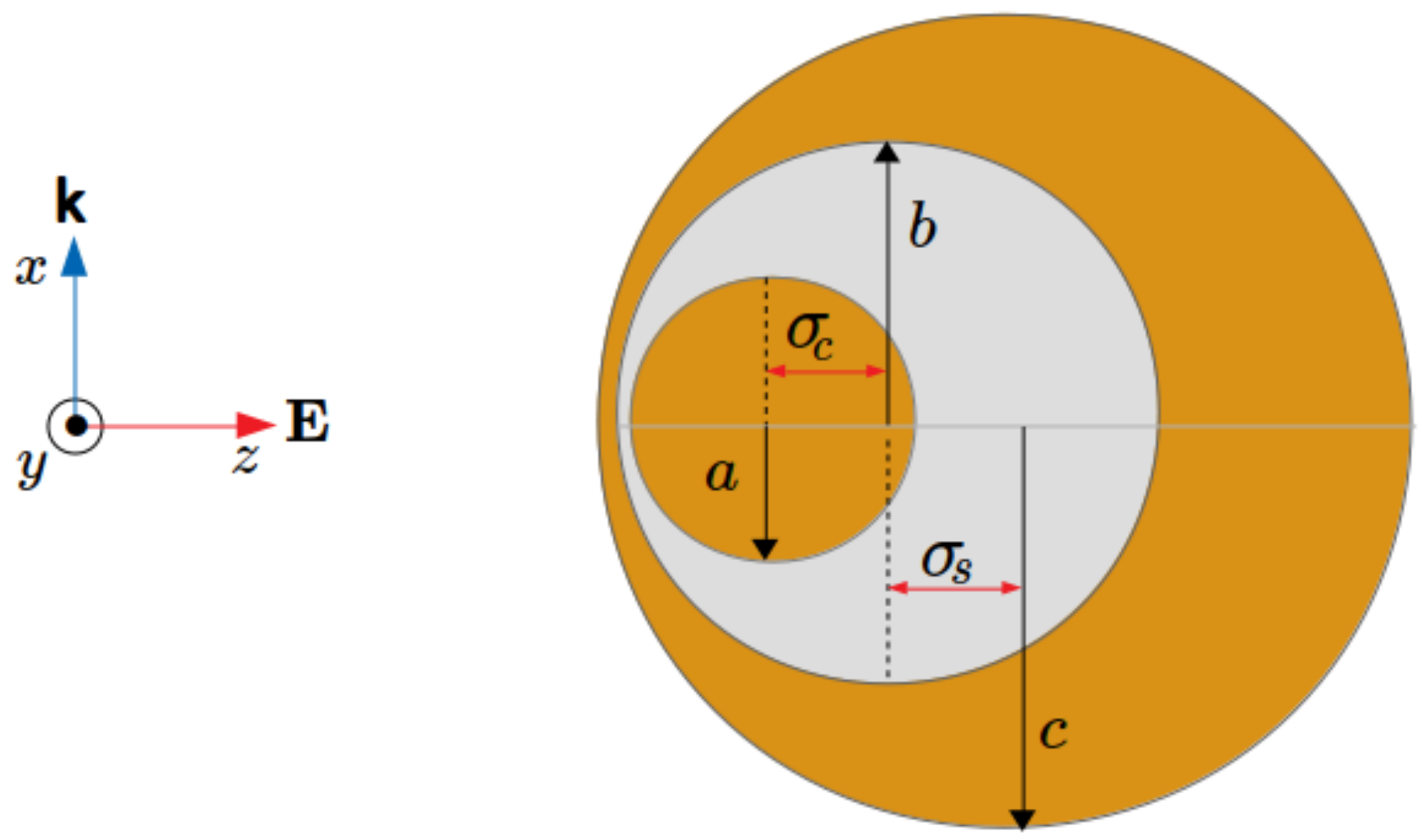}
	\caption{Model geometry of the metal-dielectric-metal MNS with a metallic core of radius $a$, an inner shell of radius $b$, giving  a dielectric layer of thickness $b-a$, and an outer shell of radius $c$, giving a metallic layer of thickness $c-b$.  An incident electric field, $\mathbf{E}$, propagating along the $x$-direction with a wavevector $\mathbf{k}$ is plane-polarized in the $z$-direction along the direction of the core and shell offsets, $\sigma_{c}$ and $\sigma_{s}$, respectively. }\label{f1}
\end{figure}

We employed the approach previously used in Ref. \cite{Nort16} where the electrostatic potentials from the solution of the Laplace equation for a sphere in a homogeneous electrostatic field are written for each region in the Fanoshell, and the respective boundary conditions inside and outside each of these regions are matched accordingly, with the application of the SHAT \cite{Cala78} at the interface where any two regions share separate centers as a result of an offset. 
This allows us to obtain the complex amplitude of the scattered dipole potential, $F_{1}$, in the medium surrounding the Fanoshell, from which the dynamic polarizability can be obtained from the expressions \cite{Moroz09,ZaZa13}: 
\begin{subequations}\label{e2}
	\begin{align}
	\alpha(\omega) & = \alpha_{s}(\omega) \left[1
	-\frac{k^{2}_{h}}{4\pi\varepsilon_{0}c}\alpha_{s}(\omega)
	-\frac{ik^{3}_{h}}{6\pi\varepsilon_{0}}\alpha_{s}(\omega)\right]^{-1}, \\
	\alpha_{s}(\omega) & = 4\pi\varepsilon_{0} c^{2}F_{1}/E_{0},
	\end{align}
\end{subequations}	
and the extinction efficiency of each Fanoshell via energy conservation \cite{BoHu08,Sam07,Kreig85,Barns16}:
\begin{equation}\label{e3}
Q_{ext} = \frac{1}{A}\left[\frac{k_{h}\Im[\alpha(\omega)]}{\varepsilon_{0}}+
\frac{k_{h}^{4}|\alpha(\omega)|^{2}}{6\pi \varepsilon^{2}_{0}}\right],
\end{equation}
where $\alpha_{s}(\omega)$ is the quasi-static dipole polarizability, $\varepsilon_{0}$ is the free-space permittivity, $A$ is the cross-sectional area of the MNS, and $k_{h} = 2\pi\sqrt{\varepsilon_{h}}/\lambda$ is the wavenumber of an incident field of amplitude $E_{0}$ and wavelength $\lambda$ propagating in a host medium of dielectric constant $\varepsilon_{h}$. 

The frequency-dependent, complex dielectric function of the metal is modelled using a Drude-Lorentz model of the form\cite{Rakic98}
\begin{equation}\label{e4}
\varepsilon(\omega) = \varepsilon_{\infty} - 
f_{0}\frac{\omega^{2}_{p}}{\omega(\omega-i\gamma_{p})}
+\sum_{j=1}^{5}f_{j}\frac{\omega^{2}_{p}}
{\omega^{2}_{j}-\omega^{2}+i\omega\gamma_{j}}, 
\end{equation}
where $\varepsilon_{c}(\omega) = \varepsilon_{s}(\omega) = \varepsilon(\omega)$ in the case of Fig. \ref{f1}, $\varepsilon_{c}(\omega)$ and $\varepsilon_{s}(\omega)$ are the dielectric constants of the metallic core and shell, respectively, $\varepsilon_{\infty}$ is the high-frequency dielectric constant of the positive ion core, $f_{0}, \omega_{p}$, and $\gamma_{p}$ are the oscillator strength, plasma frequency, and damping rate of the free electrons, respectively, $j$ is the number of Lorentz oscillators with oscillator strength $f_{j}$, frequency $\omega_{j}$, and damping rate $\gamma_{j}$ associated with bound electrons, and $\omega$ is the frequency of the incident field. The values of these parameters for different metals can be found in Ref. \cite{Rakic98}.

To find $F_{1}$, a matrix equation with $F_{l}$ unknowns is solved for $l$ multipoles, giving (see the Supporting Information \ref{ESI}): 
\begin{equation}\label{d25}
\sum_{n=1}^{N}p_{l}L_{ln}
\sum_{l=1}^{N}S_{nl}v_{l}F_{l} 
+ \sum_{n=1}^{N}q_{l}K_{ln}
\sum_{l=1}^{N}T_{nl}v_{l}F_{l} 
=  
\sum_{n=1}^{N}p_{l}L_{ln}
\sum_{l=1}^{N}U_{nl}
+\sum_{n=1}^{N}q_{l}K_{ln}
\sum_{l=1}^{N}V_{nl}, 
\end{equation}
where 	
\begin{subequations}
	\begin{align}
	S_{nl} & = \tilde{c}_{n}M_{nl}x_{l} +\tilde{d}_{n}N_{nl}, 
	T_{nl} = \tilde{a}_{n}M_{nl}x_{l} +\tilde{b}_{n}N_{nl}, \\
	U_{nl} & = \tilde{c}_{n}M_{nl}g_{l} +\tilde{d}_{n}N_{nl}u_{l}, 
	V_{nl} = \tilde{a}_{n}M_{nl}g_{l} +\tilde{b}_{n}N_{nl}u_{l}, \label{e5b}
	\end{align}
\end{subequations}

\begin{subequations}\label{d24}
	\begin{align}
	K_{ln} & = \left(\begin{array}{c}
	n\\l
	\end{array}\right)\frac{a^{l}\sigma_{c}^{n-l}}{b^{n}}
	\begin{cases}
	1,& n \ge l\\
	0,& n<l
	\end{cases},~~~
	L_{ln} = (-1)^{l-n}\left(\begin{array}{c}
	l\\n
	\end{array}\right)\frac{b^{n+1}\sigma_{c}^{l-n}}{a^{l+1}}
	\begin{cases}
	1,& l \ge n\\
	0,& l<n
	\end{cases}, \\  
	M_{nl} & = \left(\begin{array}{c}
	l\\n
	\end{array}\right)\frac{b^{n}\sigma_{s}^{l-n}}{c^{l}}
	\begin{cases}
	1,& l \ge n\\
	0,& l<n
	\end{cases},~~~
	N_{nl} = (-1)^{n-l}\left(\begin{array}{c}
	n\\l
	\end{array}\right)\frac{c^{l+1}\sigma_{s}^{n-l}}{b^{n+1}}
	\begin{cases}
	1,& n \ge l\\
	0,& n<l
	\end{cases}, 	
	\end{align}
\end{subequations}	
and 
\begin{subequations}\label{d26}
	\begin{align}
	p_{l} & = l\varepsilon_{c}(\omega)+(l+1)\varepsilon_{d},  
	q_{l} = l[\varepsilon_{c}(\omega)-\varepsilon_{d}],\\ 
	\tilde{a}_{n} & = [n\varepsilon_{s}(\omega)+(n+1)\varepsilon_{d}]/(2n+1)\varepsilon_{d},
	\tilde{b}_{n} = (n+1)[\varepsilon_{d}-\varepsilon_{s}(\omega)]/(2n+1)\varepsilon_{d},\\ 
	\tilde{c}_{n} & = n[\varepsilon_{d}-\varepsilon_{s}(\omega)]/(2n+1)\varepsilon_{d},
	\tilde{d}_{n} = [(n+1)\varepsilon_{s}(\omega)+n\varepsilon_{d}]/(2n+1)\varepsilon_{d}, \\
	v_{l} & = [l\varepsilon_{s}(\omega)+(l+1)\varepsilon_{h}]/(2l+1)\varepsilon_{s}(\omega), 
	g_{l} = y_{l}+u_{l}x_{l}, \\
	x_{l} & =  (l+1)[l\varepsilon_{s}(\omega)-\varepsilon_{h}]/[l\varepsilon_{s}(\omega)+(l+1)\varepsilon_{h}], \\ 
	y_{l} & = E_{0}c(l+2)\delta_{1l}\varepsilon_{h}/[l\varepsilon_{s}(\omega)+(l+1)\varepsilon_{h}], 
	u_{l} = E_{0}c\delta_{1l}[l\varepsilon_{s}(\omega)-\varepsilon_{h}]/(2l+1)\varepsilon_{s}(\omega),\\
	\delta_{1l} & = 
	\begin{cases}
	1,& l=1~~~\text{(bright mode)} \label{e7g}\\
	0,& l\geq 2~~~(\text{dark modes})
	\end{cases}.
	\end{align}
\end{subequations}
Here, 
$K_{ln}$ and $L_{ln}$ are the coupling constants
between the solid plasmons of the metal core and the cavity plasmons of the dielectric shell for the same ($l = n$) and different ($l \neq n$) angular momentum numbers, respectively, $M_{nl}$ and $N_{nl}$ are the coupling constants 
between the solid plasmons of the metallic shell and the cavity plasmons of the dielectric shell for the same 
($n = l$) and different ($n \neq l$) angular momentum numbers, respectively, and $\varepsilon_{d}$ is the dielectric constant of the dielectric shell. 

The simulations were performed in the RF module of COMSOL Multiphysics using a spherically-symmetric perfectly-matched layer (PML) and scattering boundary conditions applied in the internal PML surface. A $z$-polarized incident plane wave, $\mathbf{E} = E_{0}e^{-ik_{h}x}\hat{e}_{z}$, propagating in the $x$-direction was applied to the MNS. The extinction efficiency of each Fanoshell was calculated from energy conservation through the following expression \cite{Grand19}:
\begin{equation}\label{e27}
Q_{ext} = \frac{1}{AP_{0}} \left[ \iiint P_{dis}dV + \iint \Re[S_{sca}]dS \right]. 
\end{equation}
The first term in Eq. \eqref{e27} is a volume integral of the total power dissipation density of the nanoparticle, $P_{dis}$. The second term is a surface integral of the real part
of the flux of the complex Poynting vector of the scattered field, $S_{sca}$, and $P_{0}$ is the power density of the incident field \cite{Grand19}.

\section{Results and discussion}
\begin{figure}[ht!]
	\centering 
	\includegraphics[width = 1.\textwidth]{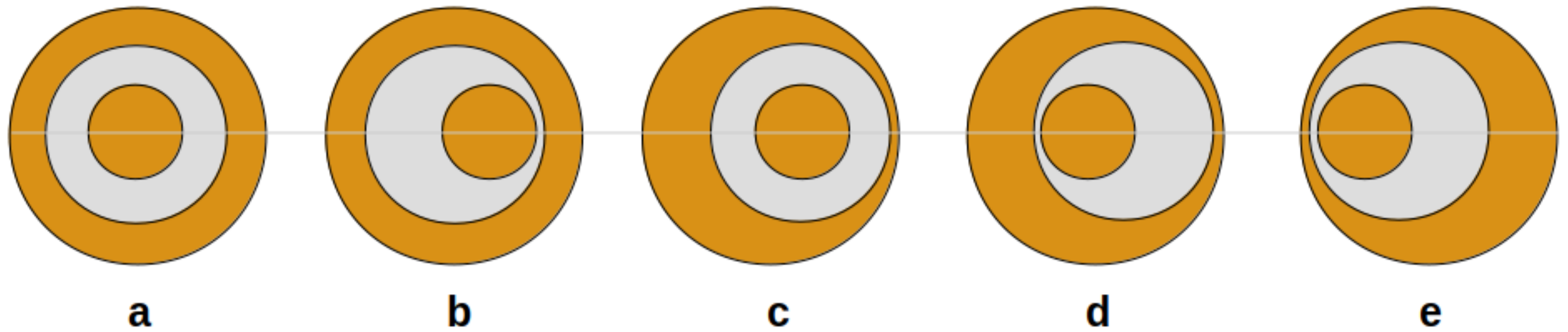}
	\caption{Fanoshells of a metal-dielectric-metal multilayer nanoshell illustrated in two dimensions.
		\textbf{a} No offset, 
		\textbf{b} Offset core, \textbf{c} Offset outer shell, \textbf{d} Offset inner shell, 
		\textbf{e} Offset core and outer shell.
	}\label{f2}
\end{figure}

We studied nanoparticle sizes within the Rayleigh regime where the electrostatic approximation is valid. The following sizes were considered: 
$a = 25$ nm, $b = 35$ nm, $c = 45$ nm, for core and shell offsets in the range $0\leq \sigma_{c}<b-a$ and $0 \leq \sigma_{s}<c-b$, respectively. The geometry we started with in Fig. \ref{f1} that led to Eq. \eqref{d25} is in form of Fig. \ref{f2}\textbf{e}: a concurrent offset core and outer shell, i.e., the direction of the core offset is opposite that of the shell offset ($\sigma_{c} = -\sigma_{s}$). This means that for this geometry, we have to choose positive offsets in Eq. \eqref{d25} since the negative sign is already accounted for in the derivation. On the other hand, the offset inner shell geometry (Fig. \ref{f2}\textbf{d}) requires that we choose negative shell offsets in order to satisfy the condition $\sigma_{c} = \sigma_{s}$ since the core and the outer shell share the same center. The direction of the offsets does not affect the optical response of the other Fanoshell configurations. 
\begin{table}[ht!]
	\centering 
	\scalebox{0.85}{
		\begin{tabular}{cccc} 
			\hline 
			&Fanoshell&Offsets&Coupling constants \\ \hline 
			\textbf{a}&No offset 	&$\sigma_{c} = \sigma_{s} = 0$&$K_{l = n}, L_{l = n}, M_{n=l}, N_{n=l}$ \\ 
			\textbf{b}&Offset core	&  $0 \leq \sigma_{c} < b-a, \sigma_{s} = 0$& $K_{ln}, L_{ln}, M_{n=l}, N_{n=l}$ \\ 
			\textbf{c}&Offset outer shell	&  $\sigma_{c} = 0, 0 \leq \sigma_{s} < c-b$& $K_{l = n}, L_{l = n}, M_{nl}, N_{nl}$ \\ 
			\textbf{d}&Offset inner shell &  $\sigma_{s} = -\sigma_{c}, 0 \leq \sigma_{s} < c-b$ & $K_{ln}, L_{ln}, M_{nl}, N_{nl}$ \\ 
			\textbf{e}&Offset core and outer shell & $0 \leq \sigma_{c} = \sigma_{s} < c-b$ & $K_{ln}, L_{ln}, M_{nl}, N_{nl}$ \\ \hline 
		\end{tabular}
	}
	\caption{Offsets and coupling constants for the realization of the optical response of each Fanoshell configuration from Eq. \eqref{d25}.
	}\label{t1}
\end{table}

All the combinations of no offset or a finite offset lead to the Fanoshell configurations shown in Fig. \ref{f2}. Table \ref{t1} lists for each of these configurations the different values of the offsets and the coupling constants associated with their core-shell plasmons. 
We solved Eq. \eqref{d25} for the first ten multipoles ($N = 10$), i.e., for $l, n = 1,2, ..., 10$, using a Python-based code, with $\varepsilon_{d} = 2.04$ and $\varepsilon_{h} = 1.77$ as the dielectric constants of silica and water (the host medium), respectively. Thus, 
Eq. \eqref{d25} is a generalized matrix equation, which when combined with Eqs. \eqref{e2} and \eqref{e3}, predicts the theoretical extinction spectra of metal-dielectric-metal multilayer Fanoshells once the dielectric constants $\varepsilon_{c}(\omega), \varepsilon_{s}(\omega), \varepsilon_{d}, \varepsilon_{h}$, the radii $a, b, c$, and the offsets $\sigma_{c}$ and $\sigma_{s}$ are given. 
We used Au-silica-Au Fanoshells to validate the theoretical model by comparing it with simulation results. This enabled us to draw comparisons with existing literature on Au-silica-Au Fanoshells \cite{Hu08,Hu10,Hala10,Jun14,Qian15}.

\subsection{Mode Suppression and Enhancement}\label{s3.1}  
\begin{figure}[ht!]
	\centering 
	\includegraphics[width = .478\textwidth]{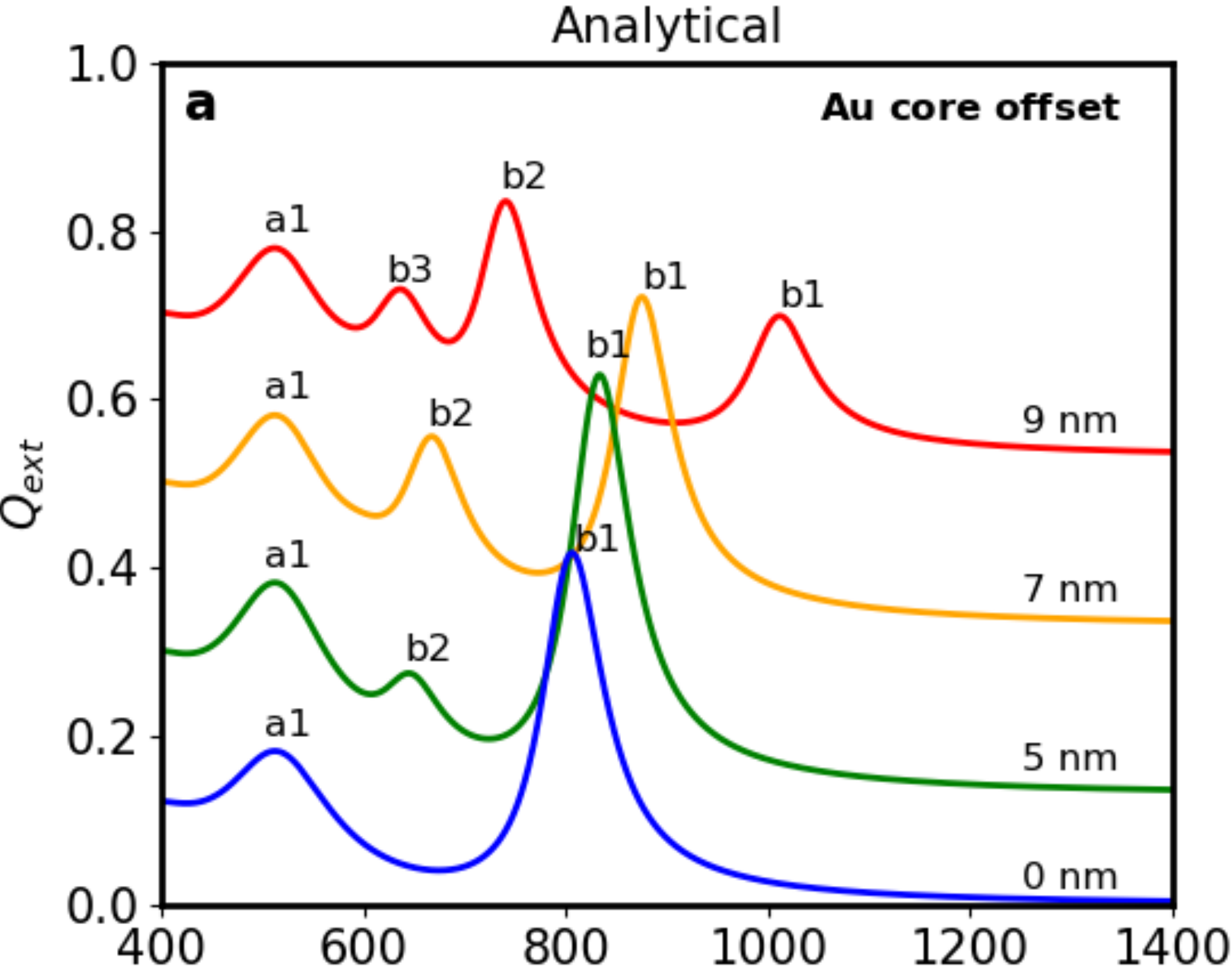}~
	\includegraphics[width = .452\textwidth]{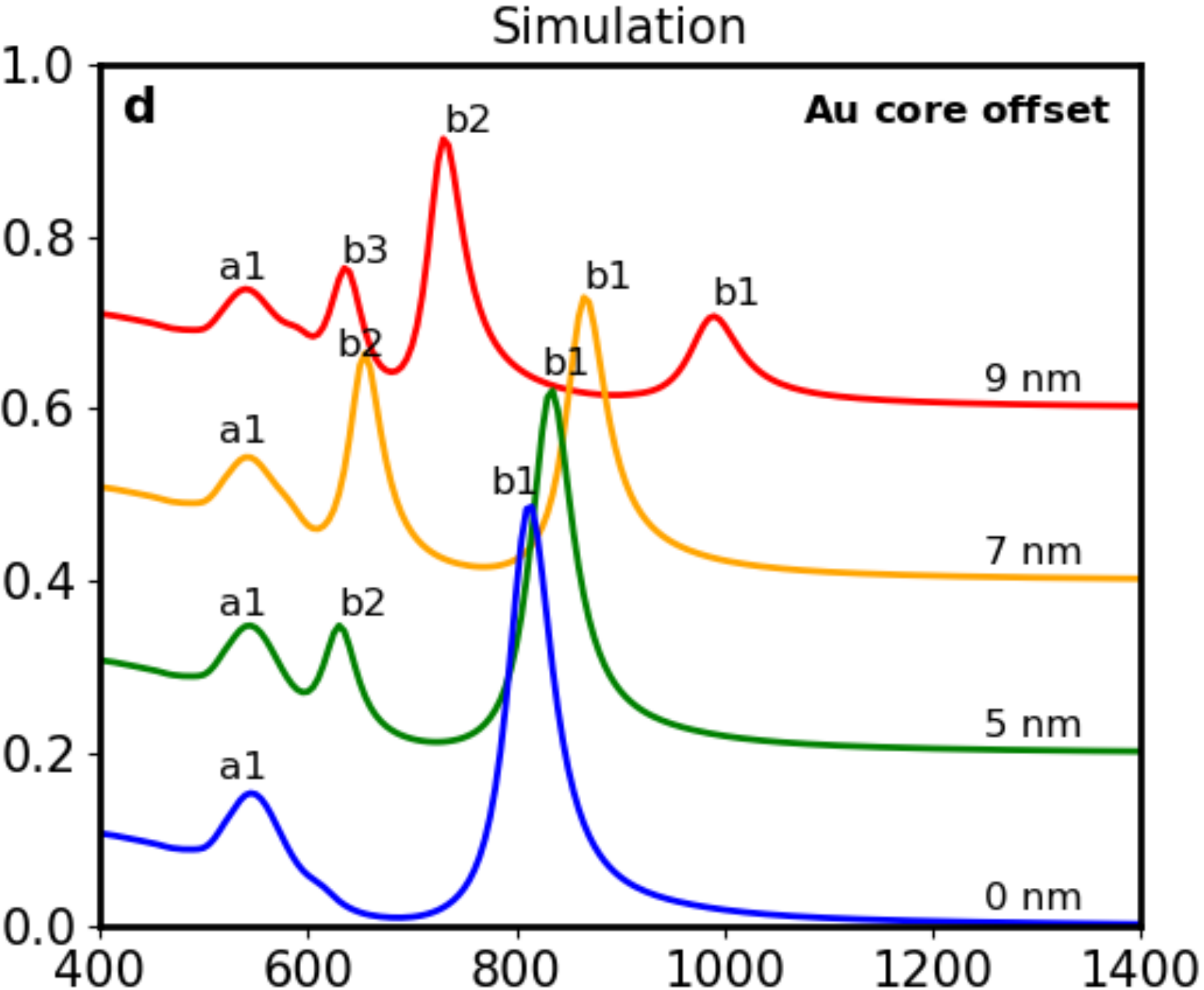}\vspace{0.2cm}\\
	\includegraphics[width = .478\textwidth]{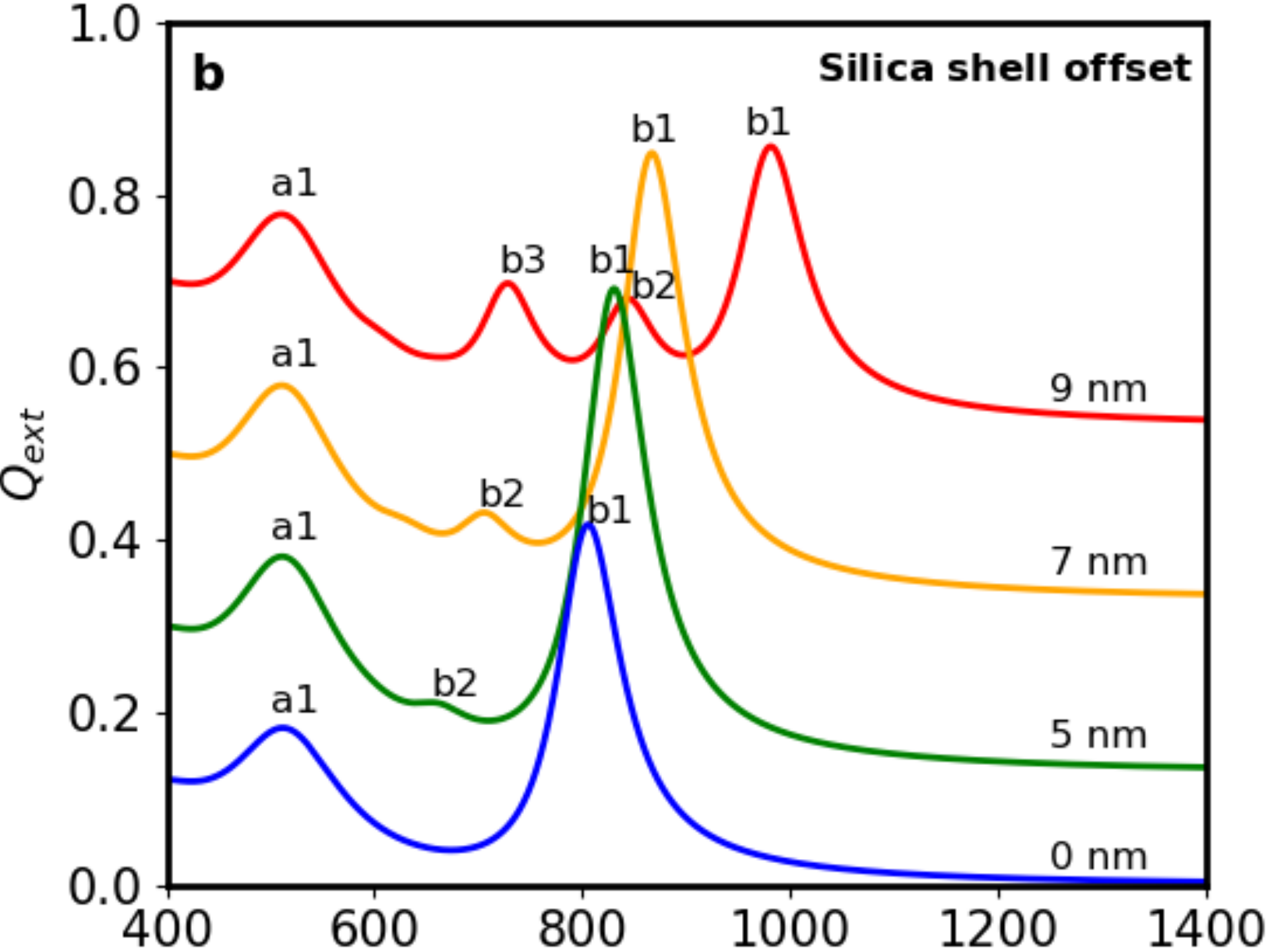}~
	\includegraphics[width = .452\textwidth]{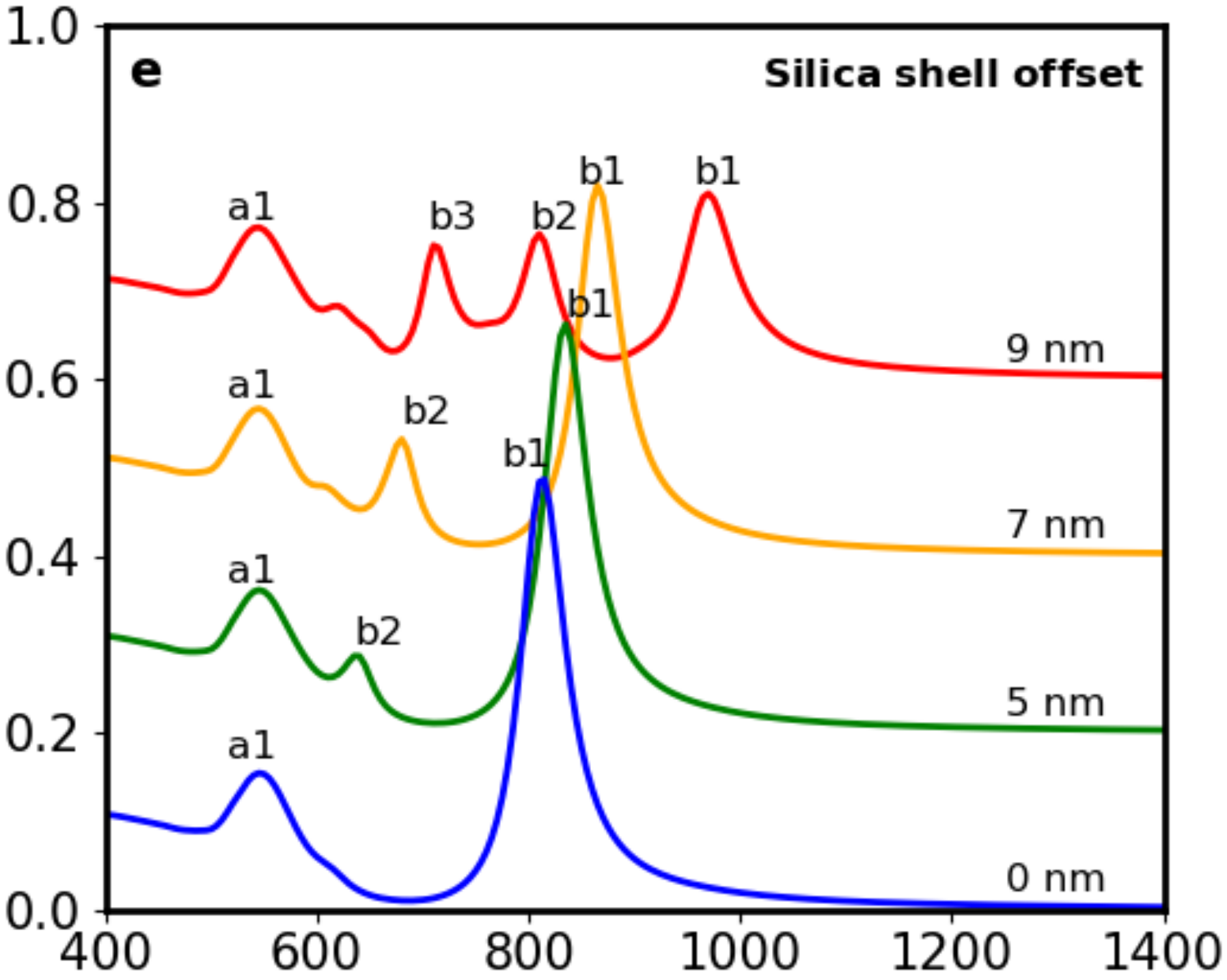}\vspace{0.2cm}\\
	\includegraphics[width = .478\textwidth]{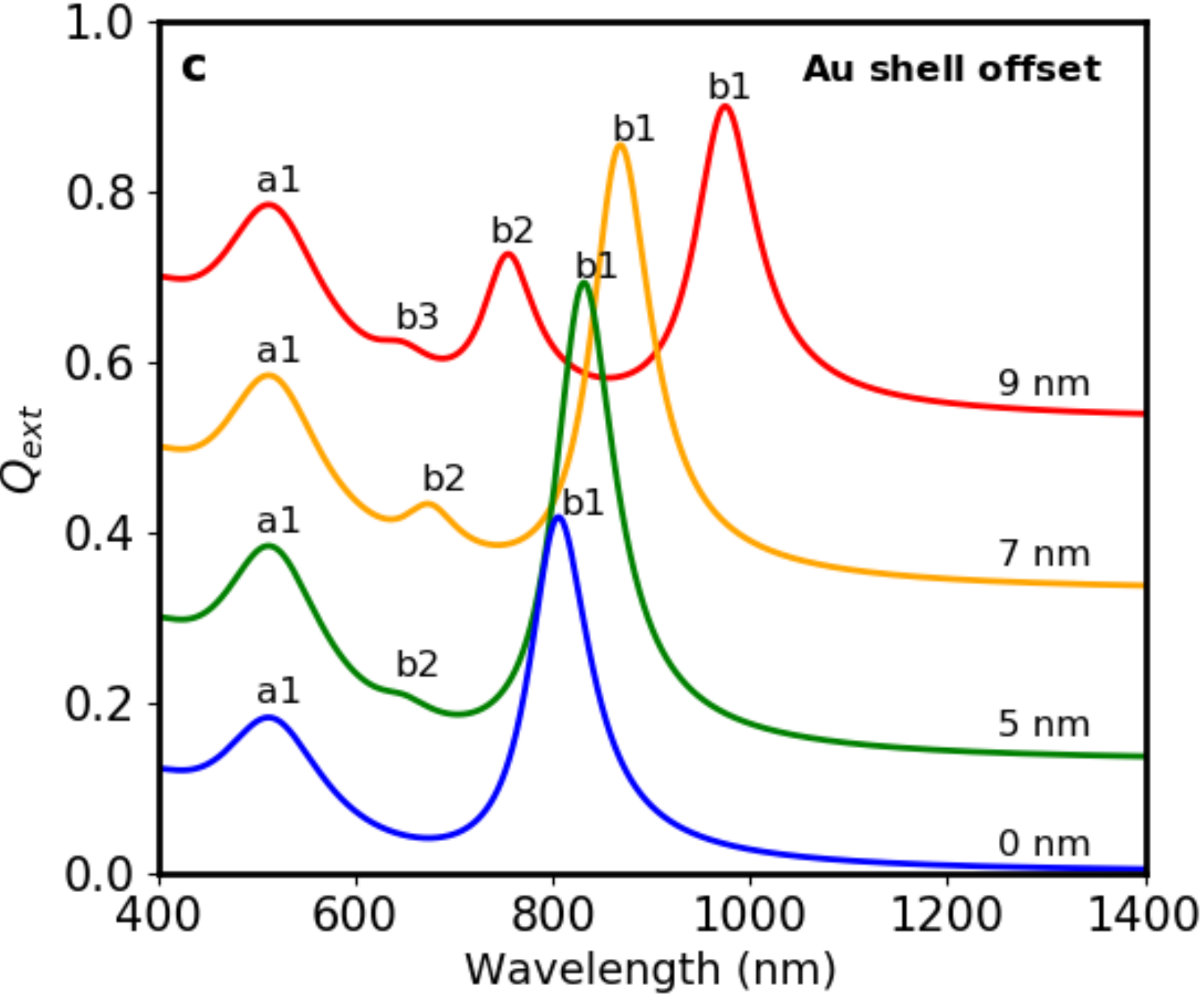}~
	\includegraphics[width = .452\textwidth]{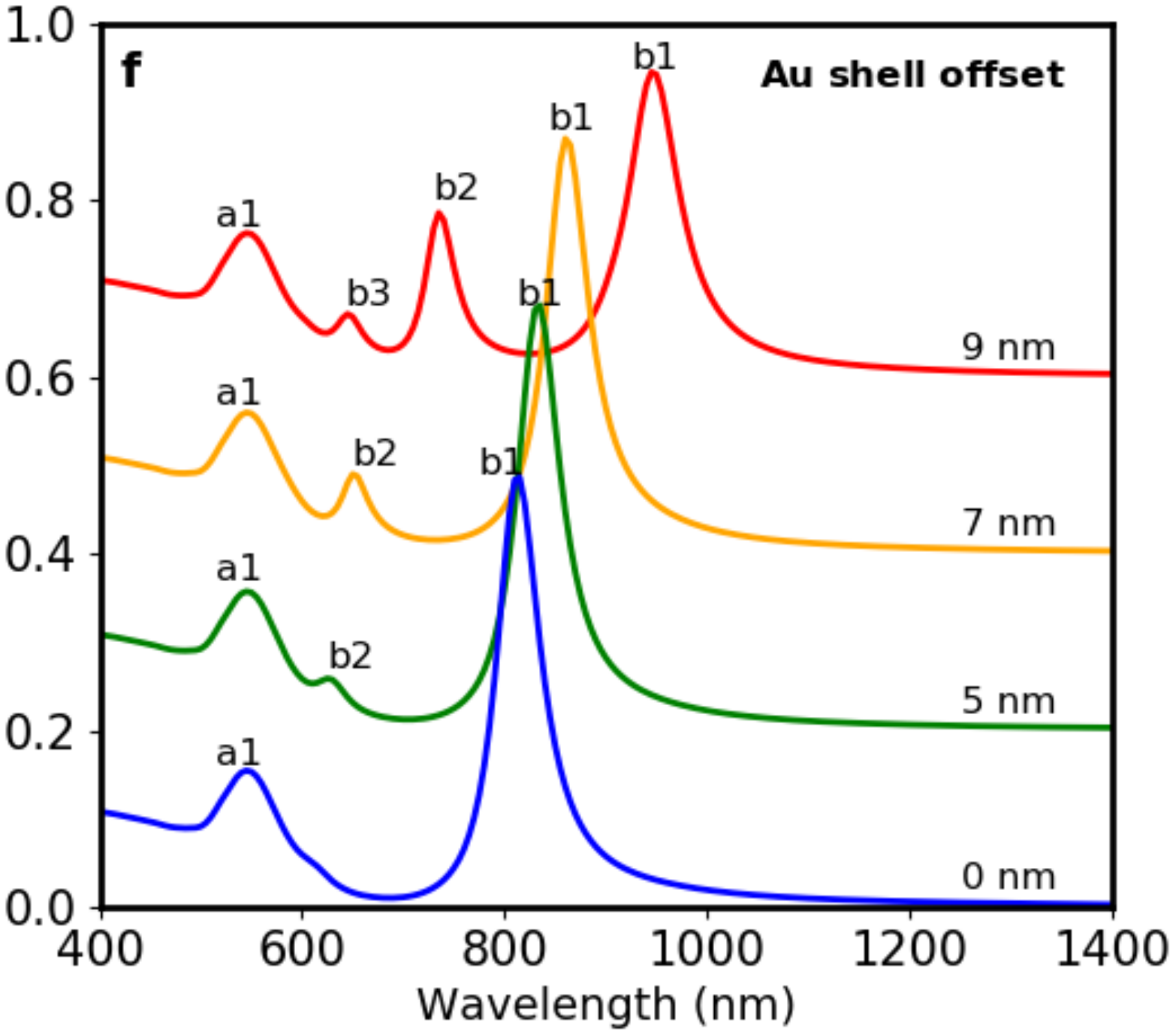}	
	\caption{Normalized extinction efficiency of the single symmetry-broken Fanoshells for the  \textbf{a}-\textbf{c} analytical results and \textbf{d}-\textbf{f} simulation results. The spectra for 5-nm, 7-nm, and 9-nm offsets have been vertically shifted by equal amounts.
	}\label{f3}
\end{figure}

\begin{figure}[ht!]
	\centering 
	\includegraphics[width = .485\textwidth]{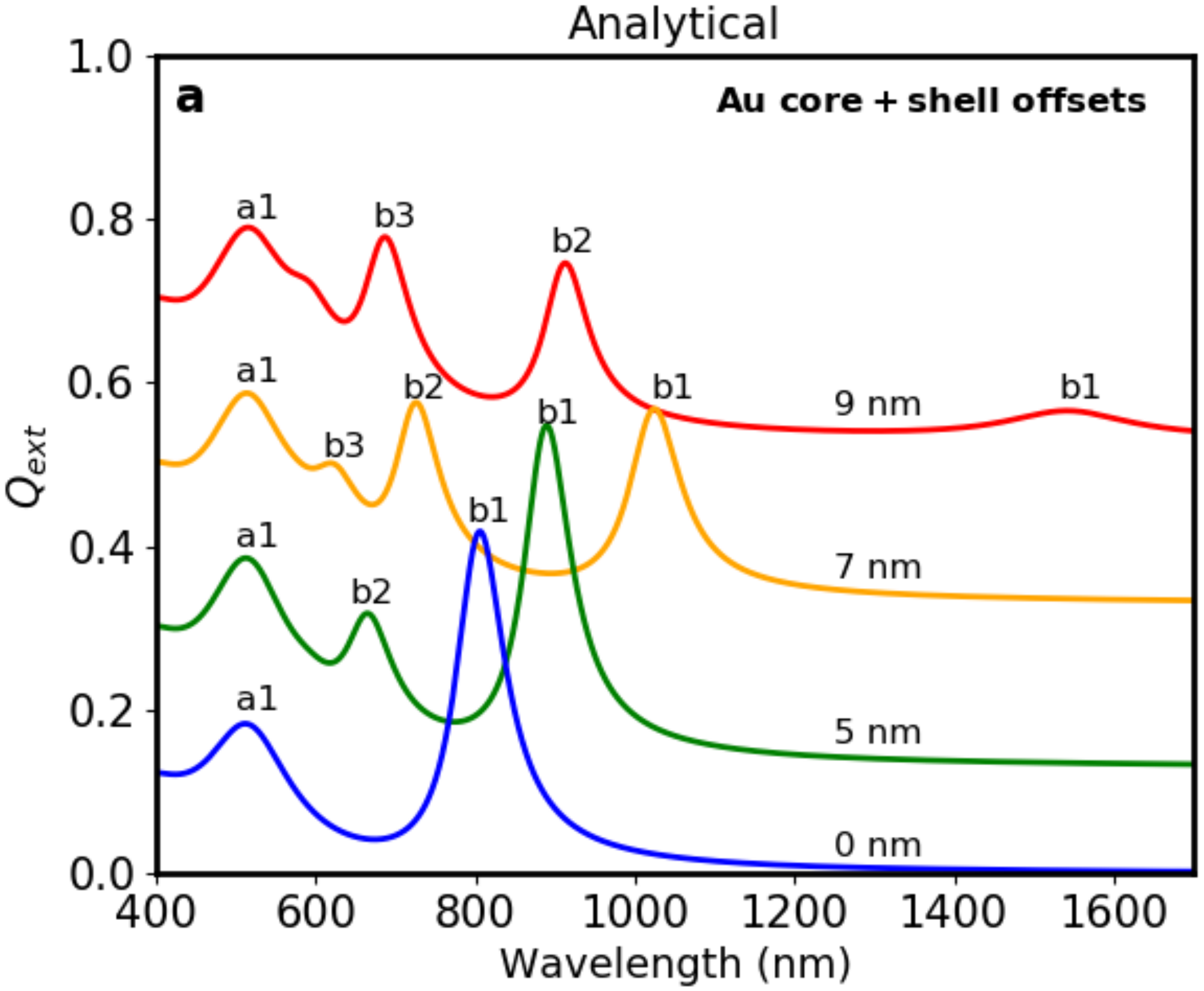}~
	\includegraphics[width = .48\textwidth]{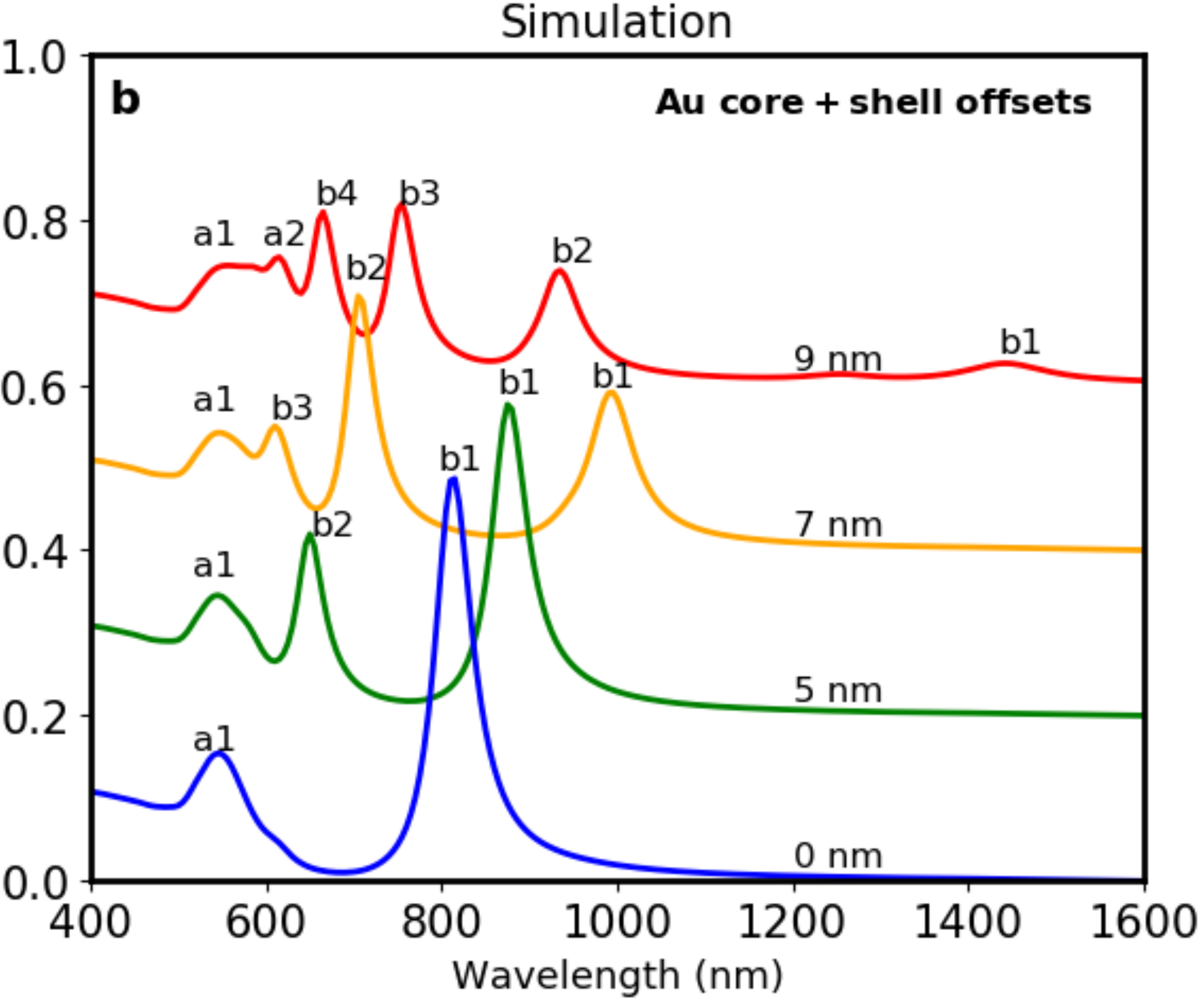}
	\caption{Normalized extinction efficiency of the dual symmetry-broken Fanoshell for the 
		\textbf{a} analytical results, and  \textbf{b} simulation results. The spectra for 5-nm, 7-nm, and 9-nm offsets have been vertically shifted by equal amounts.
	}\label{f4} 
\end{figure}

We denote the anti-bonding and bonding dipolar LSPRs by $a1$ and $b1$, respectively. Let $1, 2, 3, ...,$ denote the dipole, quadrupole, octupole modes, and so on, respectively. 
As shown in Fig. \ref{f3} and Fig. \ref{f4}, the Fanoshell with no offset (blue curves) does not support dipole-active,  multipolar  LSPRs. 
Since the coupling constants are independent of the offsets (Table \ref{t1}, \textbf{a}), the Fano effect fails to occur. Ref. \cite{Hu10} has also shown that multipolar LSPRs are dark at offsets small compared to the shell thickness.
The reason for this is that the offset-dependent coupling constants in Eq. \eqref{d24} are weak at small offsets. Thus, we used large offsets in this section to ensure that the multipolar peaks are visible as a result of strong coupling of the bright mode to the dark modes of the MNS. 

In the extinction spectra of the single symmetry-broken Fanoshells shown in Fig. \ref{f3}, 
the theoretical model (Fig. \ref{f3}(\textbf{a}-\textbf{c})) agrees well with the simulation results (Fig. \ref{f3}(\textbf{d}-\textbf{f})).
The number and positions of the LSPRs predicted by both methods are very similar. In this work, we are interested in the LSPRs, especially $b1$. 
The anti-bonding dipole LSPR, $a1$, indexed around 535 nm in Fig. \ref{f3}(\textbf{a}-\textbf{c}) and around 545 nm in Fig. \ref{f3}(\textbf{d}-\textbf{f}),
does not undergo any wavelength shift with a change in the offset. This is because in these Fanoshells, $a1$ does not undergo mode suppression.
The inability of $a1$ to undergo mode suppression and a subsequent wavelength shift in single symmetry-broken Fanoshells is probably because neither the core-offset-dependent coupling constants, $K_{l\neq n}$ and $L_{l\neq n}$, nor the shell-offset-dependent coupling constants, $M_{n\neq l}$ and $N_{n\neq l}$, are sufficient to induce the coupling of $a1$ to higher-order anti-bonding resonances. The anti-bonding mode is formed by the hybridization of the sphere plasmons of the Au core and the anti-bonding mode of the core-shell plasmons of the silica--Au nanoshell, while the bonding mode
is formed by the hybridization of the sphere plasmons of the Au core and the bonding mode of the core-shell plasmons of the silica--Au nanoshell \cite{Hu10,Zhu11,Hu08}. For the size of the MNS geometry considered in this study, and other similar sizes where $b-a = c-b$, $a1$ is usually weak compared to the bonding dipole LSPR, $b1$, since the LSPR of the anti-bonding core-shell plasmons is also weak \cite{Hu08,Hu10,Jun14}. 
Thus, such sizes are chosen in order to preferentially enhance $b1$, since it has been shown to contribute significantly to the formation of dipole-active, multipolar LSPRs \cite{Khan11,Hu10,Qian15}, compared to geometry sizes where the enhancement of $a1$ was preferred \cite{Qian15}.  

As shown in Figs. \ref{f3} and \ref{f4}, the bonding LSPRs --- $b1, b2$, and $b3$ --- dominate the extinction spectra of the Fanoshells, in agreement with previous literature \cite{Hu10,Qian15,Khan11,Ho10}. 
Fig. \ref{f3} shows that as the offset approaches the shell thickness, $b2$ and $b3$ become 
more enhanced and $b1$ becomes more suppressed. This is due to an increase in the coupling constants. The Fanoshell with a Au core offset (Figs. \ref{f3}\textbf{a} and  \ref{f3}\textbf{d}) supports a significant suppression of $b1$ and enhancement of $b2$ and $b3$ followed by the Fanoshell with a silica shell offset (Figs. \ref{f3}\textbf{b} and  \ref{f3}\textbf{e}) and the Fanoshell with a Au shell offset (Figs. \ref{f3}\textbf{c} and \ref{f3}\textbf{f}). Due to this reason, the Au core offset geometry supports larger offset-dependent wavelength shifts in
$b1$ compared to the other two Fanoshells. In Figs. \ref{f3}\textbf{a} and  \ref{f3}\textbf{d}, $b1$, indexed around 805 nm at 0-nm offset (blue curve), shifts to $\sim$1010 nm when the core is offset at 9-nm (red curve). This is more than the wavelength shifts in the silica shell offset where $b1$ shifts from $\sim$805 nm at 0-nm offset to $\sim$983 nm at 9-nm offset  (Figs. \ref{f3}\textbf{b} and \ref{f3}\textbf{e}) and in the Au shell offset where it shifts from $\sim$805 nm at 0-nm offset to $\sim$976 nm at 9-nm offset (Figs. \ref{f3}\textbf{c} and \ref{f3}\textbf{f}). 

On the other hand, the bonding multipolar LSPRs undergo larger offset-dependent wavelength shifts in Fanoshells with an offset shell. For instance, in Figs. \ref{f3}\textbf{b} and \ref{f3}\textbf{e} (silica shell offset), $b2$ shifts from $\sim$658 nm at 5-nm offset (green curve) to $\sim$843 nm at 9-nm offset, and from $\sim$645 nm to $\sim$754 nm for the same offsets in the spectra of the 
Au shell offset (Figs. \ref{f3}\textbf{c} and \ref{f3}\textbf{f}). These shifts are slightly larger than the wavelength shift in 
the spectra of Au core offset where $b2$ shifts from $\sim$645 nm at 5-nm offset to $\sim$743 nm at 9-nm offset (Figs. \ref{f3}\textbf{a} and \ref{f3}\textbf{d}), because $K_{ln}$ and $L_{ln}$ contribute more to the enhancement of the multipolar LSPRs and wavelength shifts in the dipolar LSPR but less to the wavelength shifts in the multipolar LSPRs compared to $M_{nl}$ and $N_{nl}$. In the former, all the coupling terms contribute to the extinction of the incident radiation (governed by the RHS of Eq. \eqref{d25}) while in the latter, only terms with $l = 1$ (the bright mode) contribute to the extinction, due to Eqs.  \eqref{e5b} and \eqref{e7g}. 
These Fanoshells can also support an octupole bonding LSPR, $b3$, at offsets near the shell thickness --- where the terms in the coupling constants contribute the most to plasmon hybridization. This is illustrated by the 9-nm offset spectra at peak positions $\sim$650 nm, $\sim$720 nm, and $\sim$680 nm in Figs. \ref{f3}\textbf{c} and \ref{f3}\textbf{f} (Au core offset), Figs. \ref{f3}\textbf{c} and  \ref{f3}\textbf{f} (silica shell offset), and  Figs. \ref{f3}\textbf{c} and  \ref{f3}\textbf{f} (Au shell offset), respectively.  

Figs. \ref{f4}\textbf{a} and \ref{f4}\textbf{b} show that the Fanoshell with concurrent Au core and shell offsets supports more enhanced multipolar LSPRs as well as stronger wavelength shifts compared to the Fanoshells discussed earlier. Here, $b1$ is nearly quenched 
for 9-nm offsets, leading to a large wavelength shift from $\sim$805 nm at 0 nm (blue curve) to $\sim$1542 nm for 9-nm offsets (red curve). Similarly, $b2$ undergoes a large shift, from
$\sim$662 nm for 5-nm offsets (green curve) to $\sim$914 nm for 9 nm offsets. As the offsets approach the shell thickness, higher-order LSPRs such as $b3, b4$ and $a2$ become dipole-active, as shown in Fig. \ref{f4}\textbf{b}. 
The absence of $a2$ and $b4$ in Fig. \ref{f4}\textbf{a} for 9-nm offsets shows that the simulation does not totally agree with the theoretical model for this Fanoshell at offsets comparable to the shell thickness. 
At this point, their indexing is not so straightforward. 
For instance, the indexing of the multipolar LSPR around 690 nm at 9-nm offsets as $b4$ in Fig. \ref{f4}\textbf{b} is a bit loose since this LSPR is probably due to the mode-mixing of $b4$ and $a2$. Similarly, at 7-nm offsets --- the onset of the suppression of $a1$ (Fig. \ref{f4}\textbf{b}) ---
$b3$, indexed around 605 nm, most likely forms due to the mode-mixing of $a2$ and $b3$. 
The anti-bonding quadrupole LSPR, $a2$, indexed around 620 nm at 9-nm offsets, though not dipole-active in Fig. \ref{f4}\textbf{a}, is visible in Fig. \ref{f4}\textbf{b} due to the mode suppression of $a1$. 
Hence, unlike single symmetry-breaking, dual symmetry-breaking can cause higher-order anti-bonding LSPRs to become dipole-active, since  
it involves the contribution of all the coupling constants, $K_{ln}, L_{ln}, M_{nl}$, and $N_{nl}$, to the Fano effect.  
Though all the coupling constants also contribute to the formation of the multipolar LSPRs in the silica shell offset geometry, the negative shell offset (Table \ref{t1}, \textbf{d}) decreases the coupling terms in $M_{nl}$ and $N_{nl}$ compared to their values with a positive shell offset (Table \ref{t1}, \textbf{e}). 

\subsection{Performance Parameters}\label{s3.2}
In this section, we will discuss and use three performance parameters --- geometrical sensitivity, spectral sensitivity, and scattering-to-absorption ratio ---  to determine the optimal Fanoshell for refractive index sensing. 
The aim of this section is not to optimize these parameters but to determine how they are affected by the Fanoshell configuration and the offset.
These parameters are also affected by the size of the MNS geometry. For instance, in Ref. \cite{Qian15}, where an outer Au shell offset was studied, it was shown that decreasing the core size enhances the sensitivity of the multipolar LSPRs. Nevertheless, core-size dependence is not investigated here.

\subsubsection{Geometrical Sensitivity}\label{ss3.2.1}
\begin{figure}[h!]
	\centering 
	\includegraphics[width = .45\textwidth]{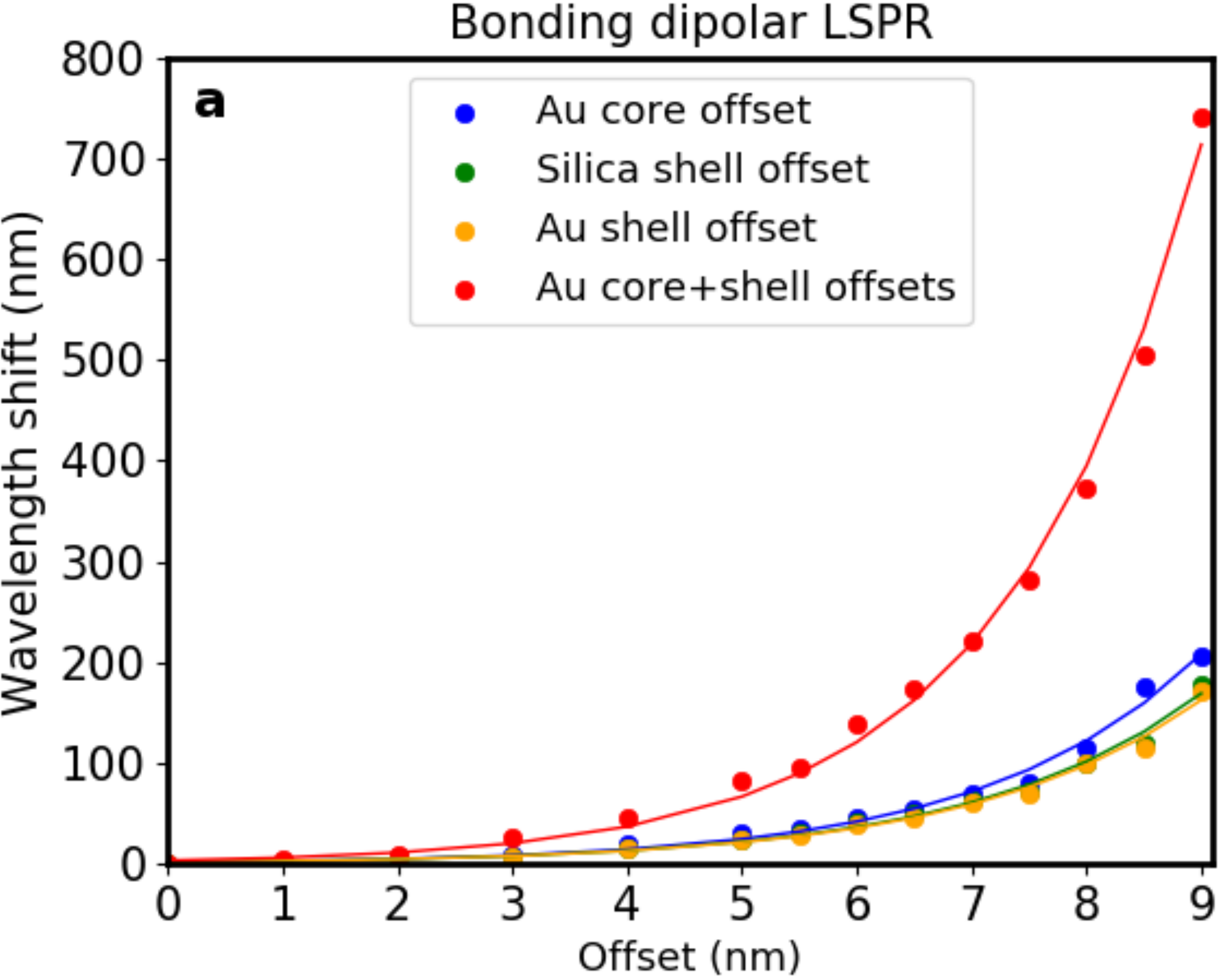}~
	\includegraphics[width = .455\textwidth]{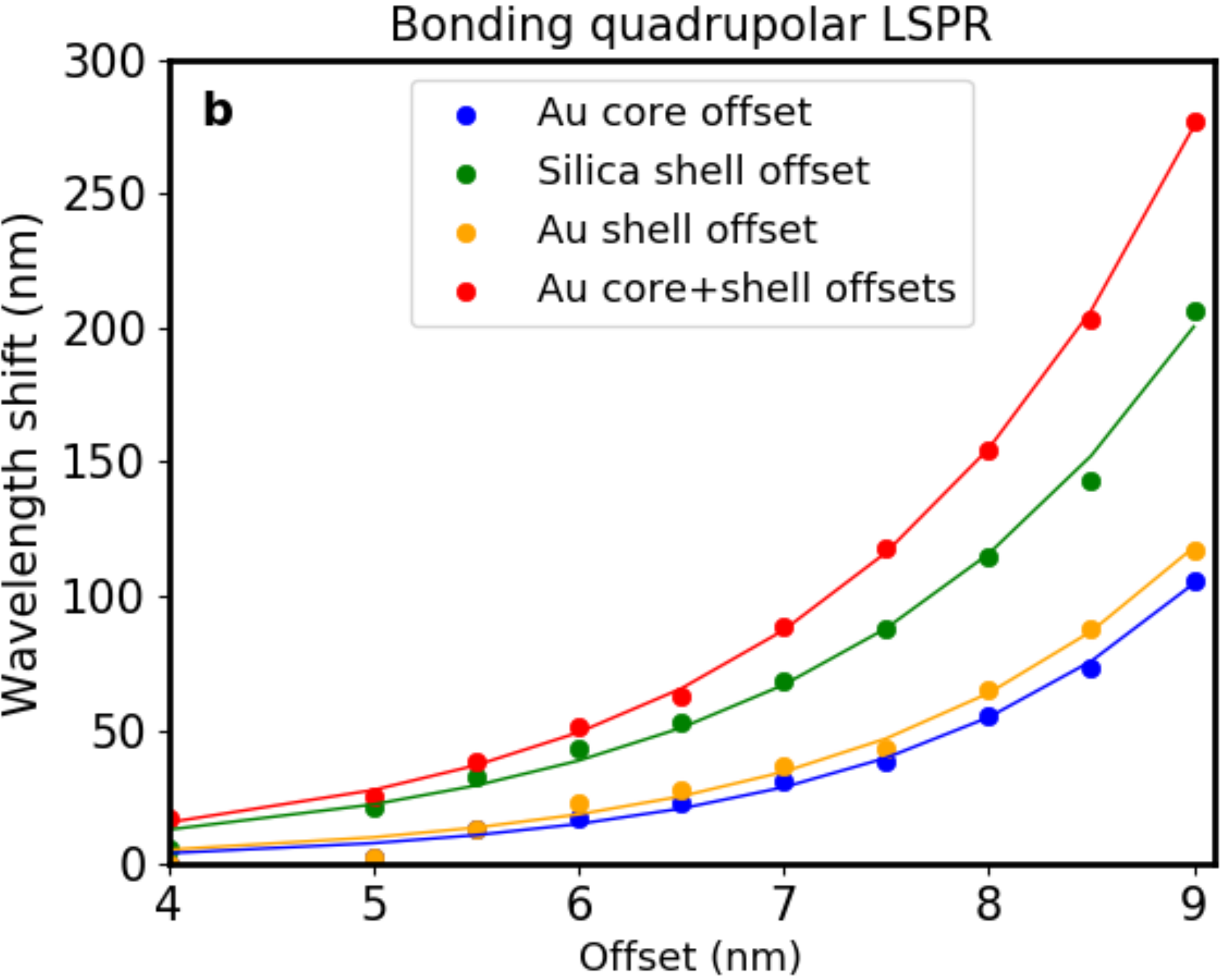}
	\caption{ Dependence of the wavelength shifts of the bonding dipolar and quadrupolar LSPRs of the Fanoshells on the offset.
		\textbf{a} Bonding dipolar LSPR, \textbf{b} Bonding quadrupolar LSPR. Dots represent calculated LSPRs and lines are exponential fits.
		The wavelength shifts in \textbf{a} were calculated using the dipolar bonding LSPR 
		of the MNS with no offset as the reference, while those in \textbf{b} were determined using the quadrupolar bonding LSPR of the MNS with a core offset at 4 nm as the reference.
	}\label{f6}
\end{figure}

In order to determine the Fanoshell configuration most sensitive to an offset, we introduce the \emph{geometrical sensitivity}. It measures the ease at which the LSPR responds to changes in the offset. A similar approach has been used in Ref. \cite{Hu08} to measure the ease at which the LSPRs respond to changes in the inner shell thickness of the no-offset Fanoshell.
Fig. \ref{f6} shows the dependence of the wavelength shifts on the offset for the two most sensitive bonding LSPRs --- the dipolar (b1) and quadrupolar (b2) LSPRs. The wavelength shift in the bonding dipolar LSPR, calculated as $b1-b1_{0}$, where $b1_{0}$ denotes the bonding dipolar LSPR of the Fanoshell with no offset, has a near-exponential growth (i.e. a redshift) with increasing offset (Fig. \ref{f6}\textbf{a}). This is a sharp contrast to the \emph{universal scaling principle} \cite{Hu08}, according to which the wavelength shift exhibits a near-exponential decay (i.e. a blueshift) with increasing thickness of the inner dielectric shell.
This inverse relationship between the universal scaling principle and offset-based, geometrical symmetry-breaking can be justified by recognizing that in the former, the coupling strength between the solid sphere plasmons of the metallic core and those of the metallic shell decreases as the thickness of the inner dielectric shell is increased, while in the latter, the coupling strength increases as the offsets approach the shell thickness.

Let $S_{g}$ denote the geometrical sensitivity or the growth rate. We can then represent the wavelength shift via an exponential function of the form 
\begin{equation}\label{27}
b1-b1_{0} = a_{0}e^{S_{g}\sigma}, 
\end{equation}
where $\sigma$ is the offset, and $a_{0}$ denotes the constant wavelength shift at small offsets in the range $0<\sigma<2$ (Fig. \ref{f6}\textbf{a}),  where the suppression of the dipolar LSPR leads to negligible shifts. 
Table \ref{t2} shows that the high values of both $S_{g}$ and $a_{0}$ in the Fanoshell with
concurrent Au core and shell offsets are responsible for the large wavelength shifts shown in 
Fig. \ref{f6}\textbf{a} (red curve). The other Fanoshells are based on single symmetry-breaking, and can only support significant wavelength shifts at large offsets.
\begin{table}[h!]
	\centering 
	\scalebox{0.9}{
		\begin{tabular}{c c c} 
			\hline 
			Fanoshell & $a_{0}$ (nm) &$S_{g}$ (nm$^{-1}$) \\ 
			\hline 
			Au core offset        & 1.72         & 0.53 \\ 
			Silica shell offset   & 1.72         & 0.51 \\
			Au shell offset       & 1.72         & 0.51 \\ 
			Au core + shell offsets & 3.46     & 0.59 \\ \hline 
		\end{tabular}
	}
	\caption{Fitting parameters obtained for the Fanoshells based on the wavelength shift (Eq. \eqref{27}), for the offset-dependent bonding dipolar LSPRs (Fig. \ref{f6}\textbf{a}). The standard errors in the exponential fits have been ignored.
	}\label{t2}
\end{table}

Fig. \ref{f6}\textbf{b} shows that the quadrupolar LSPRs are less sensitive to changes in the offset compared to the dipolar LSPRs. These LSPRs are not dipole-active at offsets that are small compared to the shell thickness, i.e., offsets in the range  $0<\sigma<4$ (Fig. \ref{f6}\textbf{b}). At those offsets, the offset-dependent coupling constants are not sufficient to cause a Fano effect. Fig. \ref{f6} also shows that the quadrupolar LSPRs are more sensitive to shell offsets than core offsets, while the dipolar LSPRs display the opposite behavior. 
This is due to the different contributions from the terms in the coupling constants (see Subsection \ref{s3.1}).

\subsubsection{Spectral Sensitivity}\label{ss3.2.2}
\begin{figure}[ht!]
	\centering 
	\includegraphics[width = .410\textwidth]{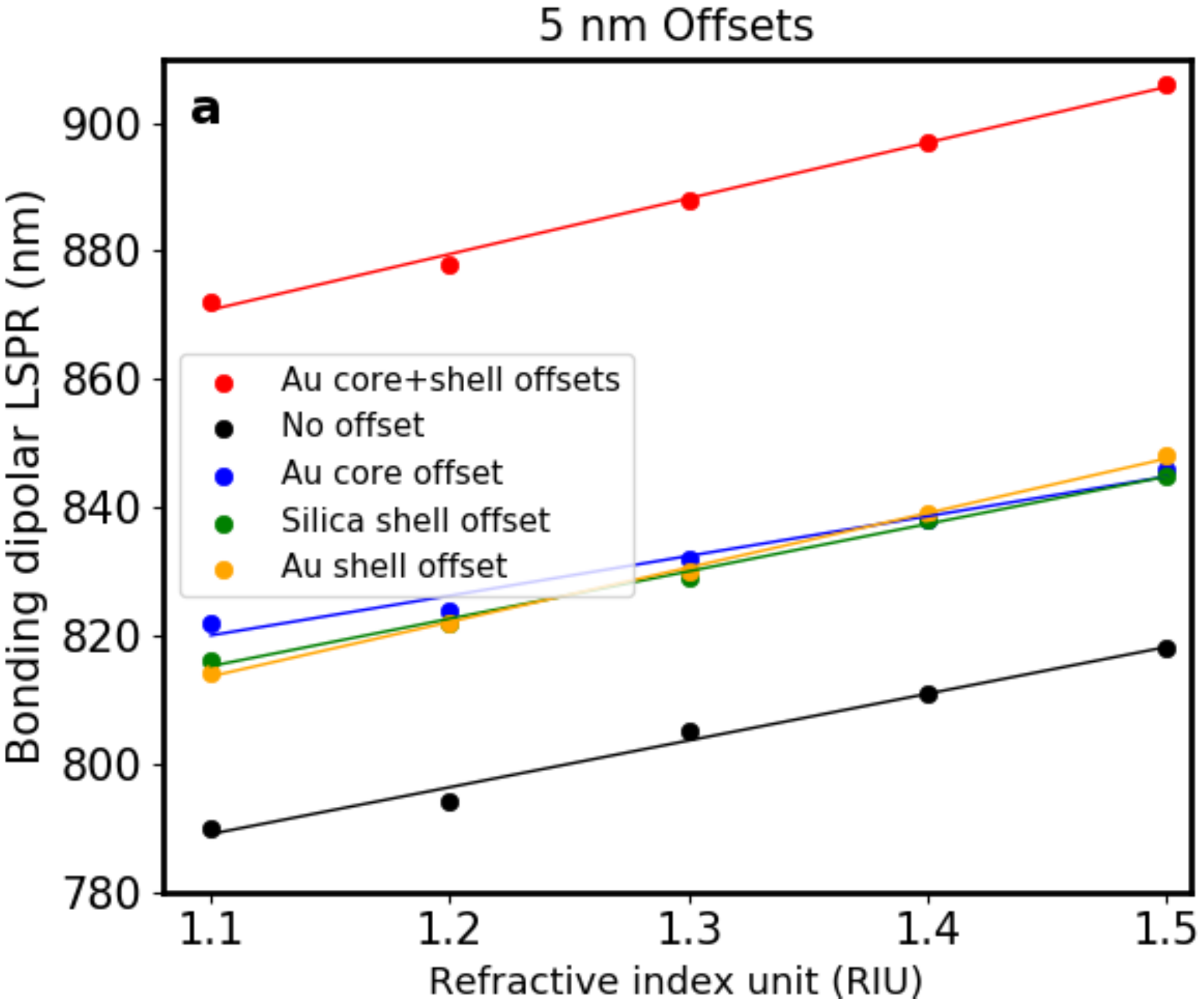}~
	\includegraphics[width = .495\textwidth]{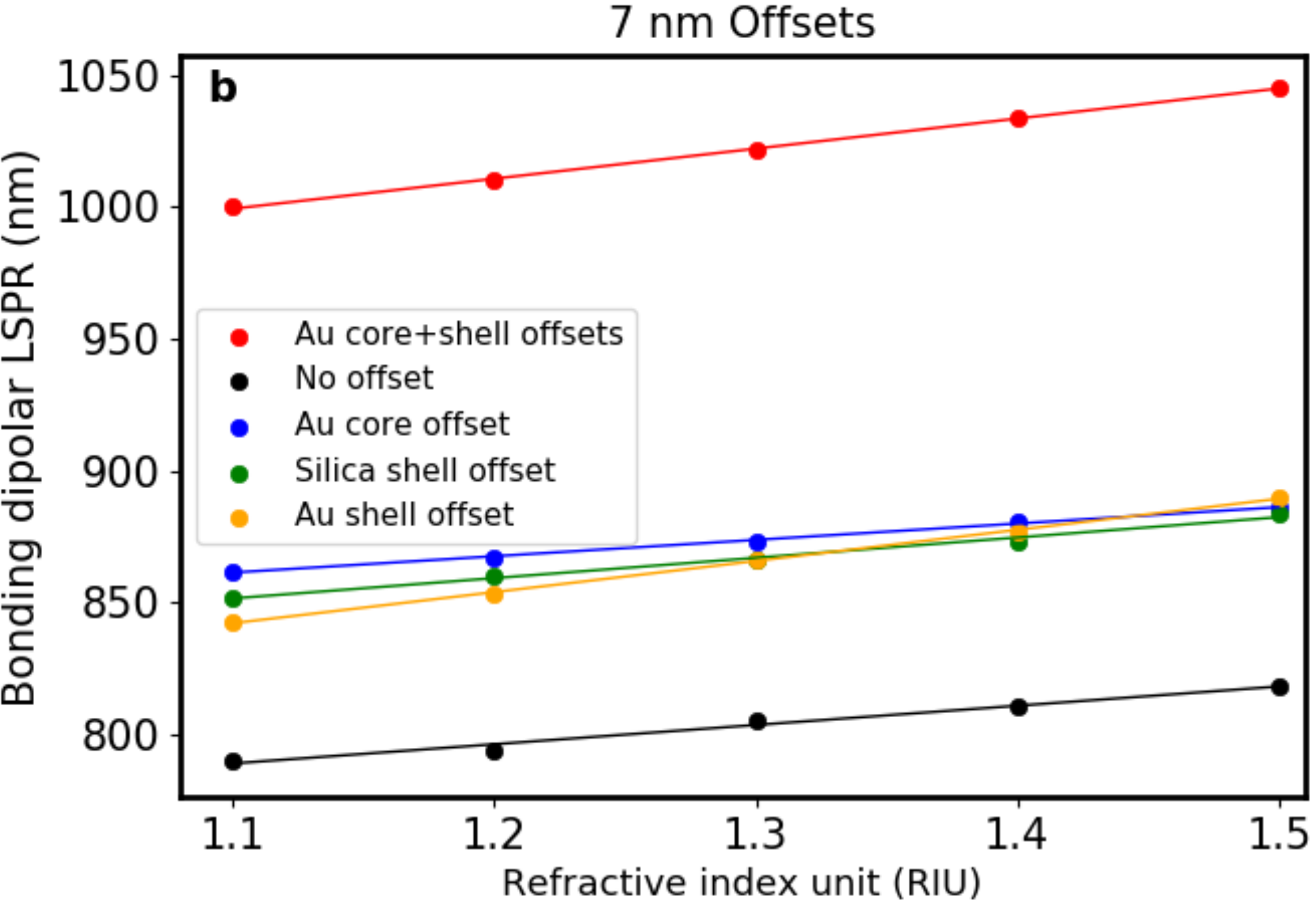}
	\caption{ Dependence of the bonding dipolar LSPRs of the Fanoshells on the refractive index of the medium at
		\textbf{a} 5 nm offsets, and \textbf{b} 7 nm offsets. Dots represent calculated LSPRs and lines are linear fits. 
	}\label{f7}
\end{figure}

The \emph{spectral sensitivity}, $S_{\lambda}$ \cite{Lee21}, also referred to as refractive index sensitivity \cite{Lee14,Soni14,Zang21}, measures the ease at which the LSPR changes with a change in the refractive index of the surrounding medium. Here, we will use $b1$, which is the most sensitive LSPR of the MNS \cite{Soni14,Ma21}, to determine the most spectrally-sensitive Fanoshell.  
As shown in Fig. \ref{f7}, $b1$ undergoes a redshift as the refractive index of the medium is increased. 
\begin{table}[ht!]
	\centering 
	\scalebox{0.8}{
		\begin{tabular}{cccc} 
			\hline 
			Fanoshell & $S_{\lambda}$ (nmRIU$^{-1}$) &$S_{\lambda}$ (nmRIU$^{-1}$)&$S_{\lambda}$ (nmRIU$^{-1}$) \\
			&  @ 5 nm offset &@ 7 nm offset&@ 7 nm offset \\ 
			&  &   &with a Ag core \\ \hline 
			No offset &    73          &             73  &  153\\ 
			Au core offset           & 62        & 63 & 125\\
			Silica shell offset   & 74        & 77  & 150\\ 
			Au shell offset       & 85        & 118 & 223\\ 
			Au core + shell offsets & 87     & 114 & 198\\ \hline 
		\end{tabular}
	}
	\caption{\small{
			Spectral sensitivities of the bonding dipolar LSPR obtained for the Fanoshells using linear fits. 
			The vertical intercepts and fitting uncertainties are not included. 
	}}\label{t4}
\end{table}
The values of $S_{\lambda}$, displayed in Table \ref{t4}, are obtained by fitting a linear function of the form $b1 = S_{\lambda}n_{h} + b0$ \cite{Soni14}, to the 
calculated values of $b1$, where $n_{h}$ is the refractive index of the host medium, and $b0$ is the vertical intercept.
At 5-nm offsets, the values of $S_{\lambda}$ are comparable for the different Fanoshells (Fig. \ref{f7}\textbf{a} and Table \ref{t4} (second column)). At 7-nm offsets, the Fanoshells with Au shell offsets become more spectrally-sensitive than the other Fanoshells (Fig. \ref{f7}\textbf{b} and Table \ref{t4} (third column)). 
Curiously, the most geometrically-sensitive Fanoshell --- the Fanoshell with concurrent Au core and shell offsets --- is not necessarily the most spectrally-sensitive.
This is likely due to two major reasons. Beside the Fanoshells being more geometrically-sensitive as the offsets approach the shell thickness, the asymmetries in Fanoshells with an outer shell offset are in direct contact with the medium while those in the core and inner shell offsets are not. 
Due to this reason, the Au core offset, which is the most geometrically-sensitive among the single symmetry-broken Fanoshells (Fig. \ref{f6}\textbf{a}), is the least spectrally-sensitive among them, as shown in Table \ref{t4}. 
Likewise, the Fanoshell with no offset is more spectrally-sensitive than the core-offset Fanoshell. 
Therefore, geometrical symmetry-breaking can be a good technique for enhancing the spectral sensitivity of MNS-based LSPR sensors only when shell offsets are involved.

\subsubsection{Scattering-to-absorption ratio}\label{ss3.2.3}
The first term in Eqs. \eqref{e3} and \eqref{e27} is the absorption efficiency, $Q_{abs}$, of the nanoparticle while the second term is the scattering efficiency, $Q_{sca}$. Let $Q_{sca}/Q_{abs}$ 
denote the scattering-to-absorption ratio. We want to determine the Fanoshell with the greatest $Q_{sca}/Q_{abs}$ at the dipole LSPRs. 
\begin{figure}[ht!]
	\centering 
	\includegraphics[width = .5\textwidth]{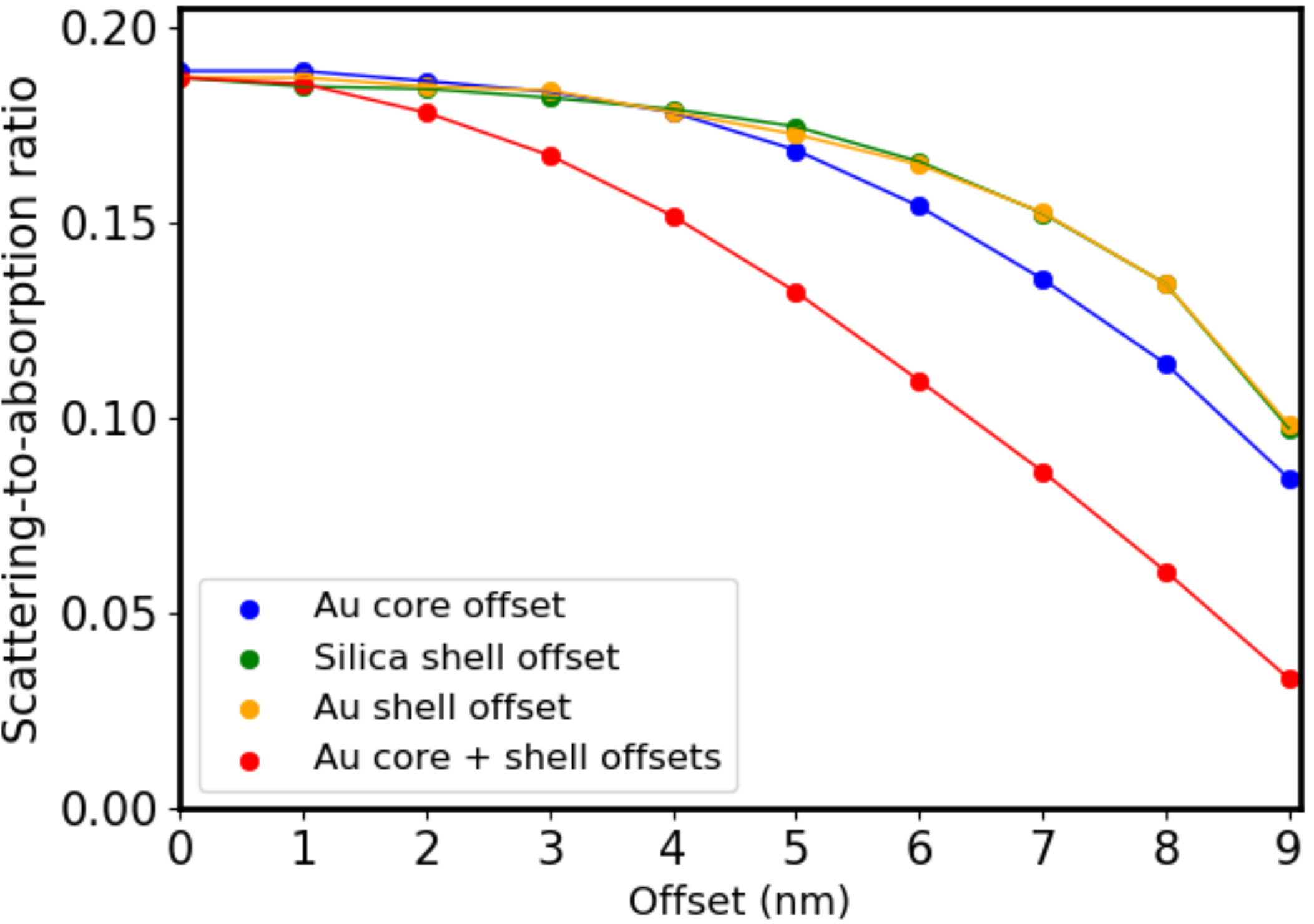}
	\caption{ Dependence of the scattering-to-absorption ratio on the offset in the Fanoshells at the dipolar LSPRs. 
		Calculated values are shown as dots connected by lines. 
	}\label{f8}
\end{figure}
Due to mode suppression, $Q_{sca}/Q_{abs}$ of the dipolar LSPRs decreases with increasing offset (Fig. \ref{f8}), in agreement with Ref. \cite{Hu10}. 

The no-offset Fanoshell (at 0-nm offset in Fig. \ref{f8}) has the largest values of $Q_{sca}/Q_{abs}$, followed by the single symmetry-broken Fanoshells. 
At large offsets, where the values of $S_{\lambda}$ are high, the Fanoshells with either inner or outer shell offsets have the largest 
values of $Q_{sca}/Q_{abs}$ since their dipolar LSPRs are less suppressed. 
Refs. \cite{Far21,Lee21} have reported that a decrease in $Q_{sca}/Q_{abs}$ is disadvantageous for sensing applications because it has a negative impact on the resolution of the sensor. Taking also the values for $S_{\lambda}$ in Table \ref{t4} into account, our results indicate that the outer shell offset is the optimal Fanoshell for sensing applications because at large offsets, its values for both $Q_{sca}/Q_{abs}$ and $S_{\lambda}$ are the highest. 

The spectral sensitivities and scattering-to-absorption ratios of these Fanoshells can be further improved by using Ag-based nanostructures \cite{Soni14,Ma21}, since the plasmon damping in Ag is less than that in Au \cite{Rakic98}. 
The last column in Table \ref{t4} shows that there is a significant improvement in $S_{\lambda}$ when the Au core is replaced with a Ag core. 
The use of other materials such as graphene \cite{Gong19} or other noble metals \cite{Kaj21,Herr16}, reduction of the core size \cite{Qian15}, or increasing the overall size of the MNS geometry \cite{Kaj21,Hu10} can enhance these parameters even further.

\section{Conclusion}
We have presented a generalized, multipole, quasi-static approach for the realization of multilayer Fanoshells by offsetting either the metallic core, the dielectric shell, the metallic shell, or both the metallic core and metallic shell. 
The equations we have derived can be used to predict the extinction spectra of metal-dielectric-metal multilayer Fanoshells of various sizes within the quasi-static regime.
Using Au-silica-Au Fanoshells, we have shown analytically that the enhancement of the multipolar modes in the extinction spectra of these Fanoshells via geometrical symmetry-breaking are accompanied by the suppression of the dipolar modes, which is in good agreement with our simulations and previous studies. Our study has revealed the nature of the coupling constants responsible for mode suppression and enhancement, as well as the optimal Fanoshell for sensing applications --- the outer shell offset. 	
The study shows that in dual symmetry-breaking of the MNS, the coupling terms involved lead to more-enhanced multipolar LSPRs, while in single symmetry-breaking, the coupling terms involved lead to less-enhanced multipolar LSPRs. 
For the Fanoshell sizes we studied, multipolar bonding LSPRs beyond the quadrupole LSPR become dipole-active when the offset approaches the shell thickness. Likewise, dual symmetry-breaking can enable the formation of visible but weak quadrupole anti-bonding LSPRs at such offsets.
We attribute this to the sensitivity of the coupling constants to large offsets. 
We also found that offset-based geometrical symmetry-breaking results in a near-exponential growth of the wavelength shifts with increasing offset, in contrast to the universal scaling principle.

\begin{acknowledgement}
	
The authors thank the National Research Foundation through Grant Nos. 120387 and 120163 for providing financial support for this work.
	
\end{acknowledgement}

\begin{suppinfo}\label{ESI}	
	
	The following files are available free of charge.
	\begin{itemize}
		\item ESI.pdf: Electronic supplementary information containing the theoretical derivations.
		\item Fanoshell.py: A Python code written for the theoretical results. 
	\end{itemize}
	
\end{suppinfo}

\bibliographystyle{unsrt}
\bibliography{manuscript.bib}
\end{document}


\maketitle
	\pagenumbering{arabic}
\section{The complex amplitude of the scattered electrostatic potential}
In order to find the complex amplitude of the scattered dipole potential described in Section 2 of the manuscript, we made
use of the orthogonality property of the Legendre function of the first kind. This is given in 
Refs. \cite{Nort16,Cala78} as: 
\begin{equation}\label{e1}
\frac{1}{\mu_{l}} \int_{-1}^{1}P_{l}(u)P_{n}(u)du = \delta_{ln}.
\end{equation}
Eq. (\ref{e1}), with $\mu_{l} \equiv 1/(l+\frac{1}{2})$, enables us to integrate the set of linear equations which were
obtained after applying the respective boundary conditions on the solution of the Laplace equation for the Fanoshell. 
This set of linear equations is then solved 
to obtain the amplitude of the scattered potential in terms of the amplitude of the incident potential. 

Let $\Phi_{h}(r_{s},\theta_{s})$ denote the electrostatic potential in the host medium. From Refs. \cite{Nort16,Luke20} we write this potential as
\begin{equation}\label{e5}
\Phi_{h}(r_{s},\theta_{s}) = \Phi_{inc}(r_{s},\theta_{s})+\Phi_{sca}(r_{s},\theta_{s}),
\end{equation}
where 
\begin{equation}  
\Phi_{inc}(r_{s},\theta_{s}) =
~ \Upsilon_{n}\Big(\frac{r_{s}}{c}\Big)^{n}P_{n}(u_{s}),~~
\Upsilon_{n} = 
\begin{cases}
-E_{0}c,& n = 1~~(\text{Bright mode})\\
0,& n \ge 2~~(\text{Dark modes})
\end{cases} \label{e6}
\end{equation}
is the incident potential, and 
\begin{equation}
\Phi_{sca}(r_{s},\theta_{s}) = \sum_{n=1}^{\infty}F_{n}\Big(\frac{c}{r_{s}}\Big)^{n+1}P_{n}(u_{s})\label{e7}
\end{equation}
is the scattered potential for an incident field polarized in the $z$-direction. Here, $r_{s}$ and $u_{s} \equiv \cos\theta_{s}$ are, respectively, the radial and angular coordinates of the metallic shell, where $r_{s}$ has been normalized by its value at the interface, $P_{n}(u_{s})$ is the Legendre polynomial of the first kind, $F_{n}$ is the complex amplitude of the scattered potential in the host medium, and $n$ is the orbital angular momentum number.

We now proceed to derive the matrix equation in $F_{n}$. 
The solution of the Laplace equation in spherical coordinates for the potentials in the metal core, the dielectric shell, the metallic shell, and in the host medium can be, respectively, represented as \cite{Nort16,Luke20}
\begin{eqnarray}
\Phi_{c}(r_{c},\theta_{c}) = \sum_{n=1}^{\infty}A_{n}\Big(\frac{r_{c}}{a}\Big)^{n}P_{n}(u_{c}),\label{d8}\\
\Phi_{d}(r_{d},\theta_{d}) = \sum_{n=1}^{\infty}\Big[B_{n}\Big(\frac{r_{d}}{b}\Big)^{n}+
C_{n}\Big(\frac{b}{r_{d}}\Big)^{n+1}\Big]P_{n}(u_{d}), \label{d9}\\
\Phi_{s}(r_{s},\theta_{s}) = \sum_{n=1}^{\infty}\Big[D_{n}\Big(\frac{r_{s}}{c}\Big)^{n}+
E_{n}\Big(\frac{c}{r_{s}}\Big)^{n+1}\Big]P_{n}(u_{s}), \label{d10}\\
\Phi_{h}(r_{s},\theta_{s}) = \Phi_{inc}(r_{s},\theta_{s})+\Phi_{sca}(r_{s},\theta_{s}), \label{d11}
\end{eqnarray}
where $r_{c}$ and $u_{c} \equiv \cos\theta_{c}$ are the radial and angular coordinates of the metal core, respectively, $r_{d}$ and $u_{d} \equiv \cos\theta_{d}$ are the radial and angular coordinates of the dielectric shell, respectively, and $P_{n}(u_{c})$ and $P_{n}(u_{d})$ are Legendre polynomials of the first kind. Note that $r_{c}$ and $r_{d}$ have been normalized by their respective values at the interface.

At the interfaces, both the potential and the normal component of the displacement field must be continuous, leading to the following boundary conditions: 
\begin{eqnarray}
\Phi_{c}\Big(r_{c},\theta_{c}\Big)\Big|_{r_{c}=a}
= \Phi_{d}\Big(r_{d},\theta_{d}\Big)\Big|_{r_{c}=a}, \label{d12}\\
\Phi_{d}\Big(r_{d},\theta_{d}\Big)\Big|_{r_{d} = b} = \Phi_{s}\Big(r_{s},\theta_{s}\Big)\Big|_{r_{d} = b}, \label{d13}\\
\Phi_{s}\Big(r_{s},\theta_{s}\Big)\Big|_{r_{s} = c} = \Phi_{h}\Big(r_{s},\theta_{s}\Big)\Big|_{r_{s} = c}, \label{d14}\\	
\varepsilon_{c}(\omega)\dfrac{\partial \Phi_{c}\Big(r_{c},\theta_{c}\Big)}{\partial r_{c}}\Big|_{r_{c} = a} = 
\varepsilon_{d}\dfrac{\partial \Phi_{d}\Big(r_{d},\theta_{d}\Big)}{\partial r_{c}}\Big|_{r_{c} = a},\label{d15}\\
\varepsilon_{d}\dfrac{\partial \Phi_{d}\Big(r_{d},\theta_{d}\Big)}{\partial r_{d}}\Big|_{r_{d} = b} = 
\varepsilon_{s}(\omega)\dfrac{\partial \Phi_{s}\Big(r_{s},\theta_{s}\Big)}{\partial r_{d}}\Big|_{r_{d} = b}, \label{d16}\\
\varepsilon_{s}(\omega)\dfrac{\partial \Phi_{s}\Big(r_{s},\theta_{s}\Big)}{\partial r_{s}}\Big|_{r_{s} = c} = 
\varepsilon_{h}\dfrac{\partial \Phi_{h}\Big(r_{s},\theta_{s}\Big)}{\partial r_{s}}\Big|_{r_{s} = c}, \label{d17}
\end{eqnarray}
where $\varepsilon_{d}$ denotes the dielectric constant of the inner shell. 

Substituting Eqs. \eqref{d8}-\eqref{d11} into Eqs. \eqref{d12}-\eqref{d17}, followed by multiplying 
Eqs. \eqref{d12} and \eqref{d15} by $P_{l}(u_{c})$, Eqs. \eqref{d13} and \eqref{d16} by $P_{l}(u_{d})$, and Eqs. \eqref{d14} and \eqref{d17} by $P_{l}(u_{s})$, 
applying the solid-harmonic addition theorem \cite{Cala78,Nort16} to the RHS of Eqs. \eqref{d12}, \eqref{d13}, \eqref{d15}, and 
\eqref{d16}, and integrating the resulting equations using the orthogonality property of the Legendre polynomial of the first kind (\eqref{e1}), we obtain the following equations:
\begin{eqnarray}
A_{l} = \sum_{n=1}^{N}K_{ln}B_{n}+\sum_{n=1}^{N}L_{ln}C_{n}, \label{d18}\\
B_{l}+C_{l} = \sum_{n=1}^{N}M_{ln}D_{n}+\sum_{n=1}^{N}N_{ln}E_{n}, \label{d19}\\
D_{l}+E_{l} = \Upsilon_{l}+F_{l}, \label{d20}\\
\varepsilon_{c}(\omega)lA_{l} = \varepsilon_{d}\Big[l\sum_{n=1}^{N}
K_{ln}B_{n}-(l+1)\sum_{n=1}^{N}L_{ln}C_{n}
\Big], \label{d21}\\
\varepsilon_{d}[lB_{l}-(l+1)C_{l}] = 
\varepsilon_{s}(\omega)[l\sum_{n=1}^{N}M_{ln}D_{n}-(l+1)\sum_{n=1}^{N}N_{ln}E_{n}], \label{d22}\\
\varepsilon_{s}(\omega)[lD_{l}-(l+1)E_{l}] = \varepsilon_{h}[ \Upsilon_{l}-(l+1)F_{l}], \label{d23}
\end{eqnarray}
where the summation has been truncated so some finite number $N$. Here, 
$K_{ln}$ and $L_{ln}$ are the coupling constants 
between the solid plasmons of the metal core and the cavity plasmons of the dielectric shell for the same ($l = n$) and different ($l \neq n$) angular momentum numbers, respectively, while $M_{ln}$ and $N_{ln}$ are the coupling constants 
between the solid plasmons of the metallic shell and the cavity plasmons of the dielectric shell, for the same 
($l = n$) and different ($l \neq n$) angular momentum numbers, respectively. The expressions for these coupling constants are
given in Eqs. (6a) and (6b) of the manuscript. 

Eliminating every other complex amplitude in Eqs. \eqref{d18}-\eqref{d23} except $F_{n}$, we obtain 
Eq. (4) of the manuscript, which is a matrix equation with $F_{l}$ as the unknowns. 

\bibliographystyle{unsrt}
\bibliography{manuscript.bib}